\documentclass[sigconf]{acmart}
\usepackage{tikz}

\usepackage{bm}
\usepackage{braket}
\usepackage{algorithm}
\usepackage{algpseudocode}

\usepackage{graphicx}
\usepackage{latexsym}
\usepackage{subcaption}
\usepackage{caption}
\usepackage{xcolor}

\usepackage{svg}

\newcommand{\circnum}[1]{%
\tikz[baseline=(char.base)]{
\node[shape=circle,fill=black,inner sep=1pt] (char) 
{\color{white}\small #1};}}
\AtBeginDocument{%
  }

\setcopyright{none}
\renewcommand\footnotetextcopyrightpermission[1]{}

\begin{document}
\fancyhf[HL]{}
\fancyhf[HR]{}
\title{Accelerating BP-based decoders for QLDPC Codes with Local Syndrome-Based Preprocessing}

\author{Wenxuan Fan}
\affiliation{%
  \institution{Kyushu University}
  \city{Fukuoka}\country{Japan}}
\email{wenxuan.fan@cpc.ait.kyushu-u.ac.jp}

\author{Yasunari Suzuki}
\affiliation{%
  \institution{RIKEN}
  \city{Saitama}\country{Japan}
}
\email{yasunari.suzuki@riken.jp}

\author{Gokul Subramanian Ravi}
\affiliation{%
  \institution{University of Michigan}
  \city{Ann Arbor}
  \country{United States}
}
\email{gsravi@umich.edu}

\author{Yosuke Ueno}
\affiliation{%
  \institution{RIKEN}
  \city{Saitama}\country{Japan}
}
\email{yosuke.ueno@riken.jp}

\author{Ilkwon Byun}
\affiliation{%
 \institution{Kyushu University}
 \city{Fukuoka}\country{Japan}}
\email{ilkwon@kyudai.jp}

\author{Koji Inoue}
\affiliation{%
 \institution{Kyushu University}
 \city{Fukuoka}\country{Japan}}
\email{inoue@ait.kyushu-u.ac.jp}

\author{Teruo Tanimoto}
\affiliation{%
 \institution{Kyushu University}
 \city{Fukuoka}\country{Japan}}
\email{tteruo@kyudai.jp}





\begin{abstract}
Due to the high error rate of qubits, detecting and correcting errors is essential for achieving fault-tolerant quantum computing (FTQC). Quantum low-density parity-check (QLDPC) codes are one of the most promising quantum error correction (QEC) methods due to their high encoding rates. BP (Belief Propagation)-based decoders are widely used and highly competitive for QLDPC codes because BP offers inherent parallelism and strong scalability. However, BP-based decoders still suffer from high decoding latency, a large portion of which is spent in the iterative BP stage.

In this paper, we propose a lightweight preprocessing step that utilizes local patterns in the syndrome to detect likely trivial error events and provide them as hints to BP-based decoders. These hints accelerate BP convergence and thereby reduce the overall decoding time. The proposed preprocessing step offers a broadly compatible approach to reducing the latency of BP-based QLDPC decodes. On the bivariate bicycle code $[[144,12,12]]$ at low physical error rates, our method achieves a $10\times$ speedup in decoding time for BP-OSD, and more than $2\times$ speedup for both BP-LSD and Relay-BP. Our method maintains the logical error rate when combined with BP-OSD and Relay-BP, while further achieving a significant reduction in logical error rate when combined with BP-LSD.

\end{abstract}



\maketitle

\section{Introduction}
Quantum computing is a promising technology for the future. It uses special quantum properties like entanglement, interference, and superposition, which may help solve certain problems much faster than classical computers~\cite{nielsen2002quantum, quantumc}. For example, Shor's algorithm~\cite{shor1997} can factor large numbers efficiently, and Grover's algorithm~\cite{10.1145/237814.237866} can speed up unstructured search problems.

Today, we are in the Noisy Intermediate-Scale Quantum (NISQ) era~\cite{Preskill2018quantumcomputingin}, where quantum computers have a limited number of qubits and high error rates in state preparation, gate operations, and measurements. These problems make it hard to run large-scale algorithms. Quantum error correction (QEC)~\cite{nielsen2002quantum} is necessary to achieve fault-tolerant quantum computing (FTQC).

One of the most widely studied QEC methods is the surface code~\cite{bravyi1998quantumcodeslatticeboundary, kitaev1997quantum}. 
It constructs logical qubits from many physical qubits arranged in a two-dimensional (2D) square lattice~\cite{PhysRevLett.98.190504, Kitaev_2003}. While surface codes are relatively straightforward to implement on hardware, their low encoding rate requires a large number of physical qubits to realize a single logical qubit.

To address this problem, researchers are exploring a variety of quantum codes, among which quantum low-density parity-check (QLDPC) codes~\cite{PRXQuantum.2.040101, gottesman2014faulttolerantquantumcomputationconstant} are a strong alternative. 
They are based on classical LDPC codes~\cite{1057683}, which are already used in technologies such as Ethernet~\cite{5437474} and 5G~\cite{8316763}. QLDPC codes can encode multiple logical qubits per block, and typically enable us to construct logical qubits with fewer physical qubits than surface codes.

For certain families of QLDPC codes, specialized decoders have been developed~\cite{10.1145/3564246.3585169,10.1145/3564246.3585101,article111,Single-Shot}, among which belief propagation with ordered statistics decoding (BP-OSD)~\cite{Panteleev_2021} is the most representative.
This decoder consists of two main stages. 
The first stage is belief propagation~(BP), in which the decoder exchanges and updates probabilistic ``beliefs'' between nodes in a bipartite graph to estimate likely errors. 
BP is widely used in classical error correction. 
However, due to quantum degeneracy~\cite{morris2024absorbingsetsquantumldpc} (i.e., different error patterns can produce the same syndrome), BP alone may not find a unique solution. 
To address this, the second stage, ordered statistics decoding~(OSD)~\cite{PhysRevResearch.2.043423}, is applied. 
OSD helps choose the most likely error pattern from several possible candidates, improving the overall accuracy of the decoding process.

\begin{figure}[t]
  \centering
  \includegraphics[width=\linewidth]{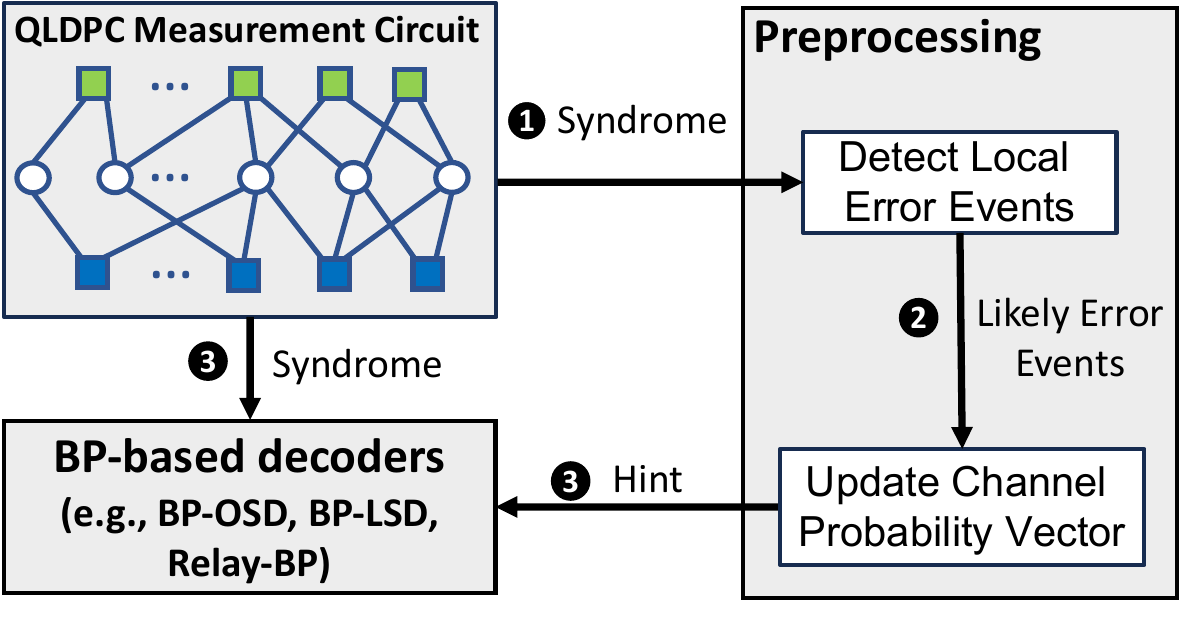}
  \caption{Workflow of the preprocessing decoder.}
  \label{fig:workflow}
\end{figure}

Although BP-OSD is a widely used high-threshold decoder for QLDPC codes, it suffers from high latency. Reducing decoding latency is crucial because the error-correction process requires real-time feedback.  High latency delays this feedback and can lead to the accumulation of errors~\cite{terhal2015quantum,holmes2020nisq,battistel2023real}. Motivated by this requirement, in Section~\ref{Motivation}, we present a comprehensive analysis of the trade-off between decoding time and logical error rate for BP-OSD. Our experimental results show that the BP stage often dominates decoding latency in practice. Importantly, this bottleneck is not specific to BP-OSD. Since BP is a shared component across a wide range of decoders, this motivates us to move beyond optimizing a single decoder instance and instead target the common bottleneck in BP-based decoding.

In Section~\ref{QPrelude}, we propose a lightweight local syndrome-based preprocessing algorithm for QLDPC codes. This preprocessing step provides informative guidance to accelerate subsequent BP-based decoders (e.g., BP-OSD, BP-LSD, and Relay-BP). We present both the software workflow and an FPGA-based hardware architecture for efficient implementation.
To the best of our knowledge, the proposed method can accelerate a wide range of BP-based decoders. The workflow of the proposed decoder is shown in Figure ~\ref{fig:workflow}. \circnum{1} Preprocessing takes the raw syndrome extracted from the QLDPC syndrome measurement circuit as input. \circnum{2} The preprocessing module analyzes the raw syndrome to detect local error events based on predefined local patterns. 
These detected events are recorded and used to update the channel probability vector. \circnum{3} The updated channel probability vector, together with the original raw syndrome, is then passed to BP-based decoders to perform final decoding.

In Section~\ref{Evaluation}, we demonstrate the broad applicability of preprocessing decoder across different platforms, noise models, and representative BP-based decoders. The results show that, at low physical error rates, our method provides substantial acceleration. For example, for the bivariate bicycle code $[[144,12,12]]$ at a physical error rate of 0.05\%, it reduces the total decoding time by 90\% when combined with BP-OSD. For relay-BP, it reduces the decoding time by more than 70\%, while maintaining the original logical error rate under the uniform circuit-level noise model.

Given the above design motivation and workflow, the key contributions of this paper can be summarized as follows:

\begin{itemize}
\item We propose a lightweight local syndrome-based preprocessing method for QLDPC codes under circuit-level noise models, including biased noise settings.

\item We develop a complete decoding pipeline and implement it in both software and FPGA-based hardware.

\item We demonstrate the broad applicability of the preprocessing across multiple noise settings and multiple representative BP-based decoding baselines.
\end{itemize}

\section{Background}
\label{background}
\subsection{Quantum Error Correction Overview}
One of the main challenges in building practical quantum computers is the high error rate of physical qubits. 
To reduce the impact of these errors on quantum computations, we use Quantum Error Correction (QEC)~\cite{nielsen2002quantum}. 
QEC works by introducing redundant qubits and by using classical devices to detect and correct errors. 
In QEC, we use check qubits and data qubits that are entangled together to protect logical qubits, which are used to execute quantum algorithms. 
Data qubits represent a logical qubit, and check qubits are used to measure the parity between different data qubits.

During QEC, we continuously measure the check qubits to gather information called syndromes. The syndromes are then sent to a classical error-estimation unit called a decoder, which estimates probable error types and locations. 
Based on this, we can apply corrections to recover the correct computational results. 
It is important to note that qubits are different from classical bits; classical bits can only suffer bit-flip errors, while qubits can suffer both bit-flip~(Pauli-$X$) and phase-flip~(Pauli-$Z$) errors. 
The Pauli-$Y$ error is a combination of both ($Y = iXZ$). 
In typical QEC schemes, we can detect and correct bit- and phase-flip errors separately.

\subsection{Quantum Low-Density Parity-Check Codes}
\begin{figure}[t]
  \centering
  \includegraphics[width=\linewidth]{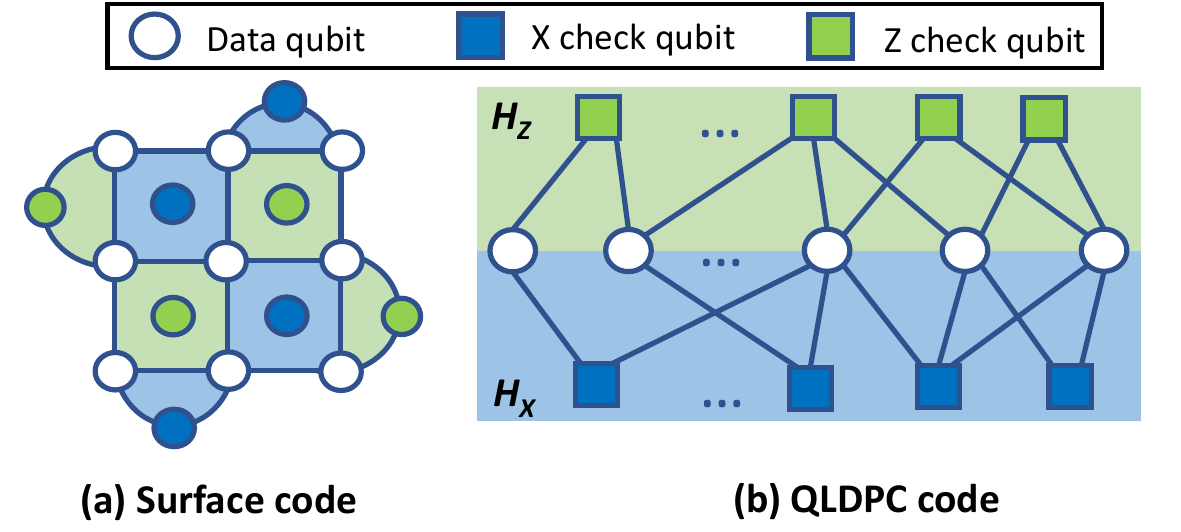}
  \caption{Structural comparison between surface codes and QLDPC codes. 
(a) Surface code. 
(b) QLDPC code.}
  \label{fig:background}
\end{figure}
QLDPC codes are a class of QEC codes that extend classical LDPC codes into the quantum domain by incorporating sparse stabilizer structures. 
Compared with surface codes (Figure~\ref{fig:background}(a)), QLDPC codes (Figure~\ref{fig:background}(b)) exhibit different connectivity structures.  Most QLDPC codes are constructed within the Calderbank-Shor-Steane (CSS)~\cite{PhysRevA.54.1098, PhysRevLett.77.793} framework. 
The CSS construction leads to two sparse binary parity-check matrices $ H_X$ and $H_Z$ , corresponding to $X$-type and $Z$-type stabilizer generators, respectively. 
These matrices must satisfy the commutativity condition $H_X\cdot H_Z^\top = 0$, which ensures that all stabilizers commute.

Due to their sparse structure and low-weight stabilizers, QLDPC codes can potentially achieve constant-rate encoding with high code-distance scaling, making them attractive for fault-tolerant quantum computing. 
In recent years, researchers have developed various families of QLDPC codes, such as HGP codes~\cite{Kovalev_2012, Tillich_2014}, 2D lattice stabilizer codes~\cite{li2024transformarbitrarygoodquantum}, generalized bicycle codes~\cite{koukoulekidis2024smallquantumcodesalgebraic}, and BB codes~\cite{bbcode}.

\subsection{BP-based Decoders}

\label{BP-based Decoders}
In surface codes, decoding problems, i.e., the task of finding the most likely error pattern from given syndromes, can be reduced to a minimum-weight perfect matching (MWPM) problem on a graph. This problem can be efficiently solved using decoders based on the blossom algorithm~\cite{Fowler_2012, Higgott_2025}. 
This conversion is enabled by a property of surface codes that each data qubit is connected by at most two check qubits.

However, we cannot apply the same strategy to QLDPC codes. 
This is because each data qubit in QLDPC codes is typically connected by more than two check qubits, and decoding problems become equivalent to MWPM problems on hypergraphs. 
Therefore, traditional decoders like MWPM are not suitable for QLDPC codes.

Instead, belief-propagation (BP)–based decoders have emerged as a promising approach for QLDPC codes decoding, including BP-OSD~\cite{Panteleev_2021}, BP-LSD~\cite{hillmann2024localizedstatisticsdecodingparallel}, Relay-BP~\cite{relay-bp}, and related variants. These BP-based decoders follow similar principles. 
As a representative example, we briefly describe BP-OSD.
The BP-OSD decoder takes three essential inputs:
(i) the measured syndrome \(\boldsymbol{\sigma}\) obtained from the QLDPC circuit;
(ii) the decoding matrix \(D \in \mathbb{F}_2^{M \times (nN_c + n)}\), where \(N_c\) is the number of measurement cycles, \(n\) is the number of check qubits, and \(M\) denotes the total number of error events considered. Each column of \(D\) corresponds to the syndrome pattern caused by a single error event in the circuit; and
(iii) the channel probability vector \(\mathbf{P} = [P_1, P_2, \dots, P_M]\), where each \(P_i\) represents the prior probability of the \(i\)-th error event.

They search the most likely low-weight error combination \(\boldsymbol{\xi}^*\in\{0,1\}^M\) satisfying 
\begin{equation}
    D\,\boldsymbol{\xi}^* = \boldsymbol{\sigma}\pmod{2},
\end{equation}
by first performing BP on the hypergraph to diffuse soft information weighted by \(\{P_i\}\), and subsequently applying OSD to rank and test candidate error vectors. The resulting \(\boldsymbol{\xi}^*\) represents the estimated set of errors that most likely occurred in the circuit.

\subsection{Related Work}

\label{Related Work}
The literature on preprocessing methods for decoding has primarily focused on surface codes. Representative examples include Promatch~\cite{10.1145/3620666.3651339}, which proposes a real-time adaptive predecoder to reduce syndrome density, and the work by Smith \textit{et al.}~\cite{Smith_2023}, which presents a local predecoder that substantially reduces the bandwidth and latency of a global decoder by performing greedy corrections. 
Pinball~\cite{pinball} further proposes a preprocessing approach for surface codes under circuit-level noise. Another example is the work by Chamberland \textit{et al.}~\cite{Chamberland_2023}, which uses local neural network decoders to reduce syndrome density before a global decoder is applied. These works demonstrate that carefully designed local preprocessing can significantly reduce the decoding latency of surface codes.

The research on QLDPC decoding has explored multiple directions to reduce decoding latency. 
One line of work replaces BP-OSD~\cite{Panteleev_2021} with alternative decoding methods, particularly machine learning–based approaches. Examples include transformer-based neural decoders~\cite{blue2025machinelearningdecodingcircuitlevel}, graph neural network decoders~\cite{ninkovic2024decodingquantumldpccodes, maan2024machinelearningmessagepassingscalable}, and neural belief propagation~\cite{PhysRevLett.122.200501}. Another direction exploits hardware acceleration and parallelization to improve runtime efficiency. Representative examples include GPU-accelerated implementations such as CUDA-QX~\cite{cudaqx}, fully parallelizable belief propagation decoders~\cite{wang2025fullyparallelizedbpdecoding}, and hardware-aware decoding frameworks such as Vegapunk~\cite{10.1145/3725843.3756084}, which introduces an online hierarchical decoding algorithm and sparse accelerator design.

In addition to alternative decoders and hardware acceleration, a large body works aim to improve the BP-based decoding framework itself. These approaches can be broadly divided into two categories.
The first category focuses on directly modifying the BP stage itself to accelerate convergence. Representative methods include Relay-BP~\cite{relay-bp} and 
SymBreak~\cite{yin2024symbreakmitigatingquantumdegeneracy}. The second category investigates post-processing techniques that operate after the BP stage, replacing the OSD step with lower-complexity algorithms. Notable examples include Ambiguity Clustering (BP+AC)~\cite{10821172}, Localized Statistics Decoding (BP+LSD)~\cite{hillmann2024localizedstatisticsdecodingparallel}, the Ordered Tanner Forest method (BP+OTF)~\cite{iolius2024almostlineartimedecodingalgorithm}, and Guided Decimation Guessing (BP+GDG)~\cite{gong2024lowlatencyiterativedecodingqldpc}.

\section{Motivation}
\label{Motivation}
\subsection{Our Key Observation}

\begin{figure}[t]
    \centering
    \begin{subfigure}{\linewidth}
        \centering
        \includegraphics[width=\linewidth]{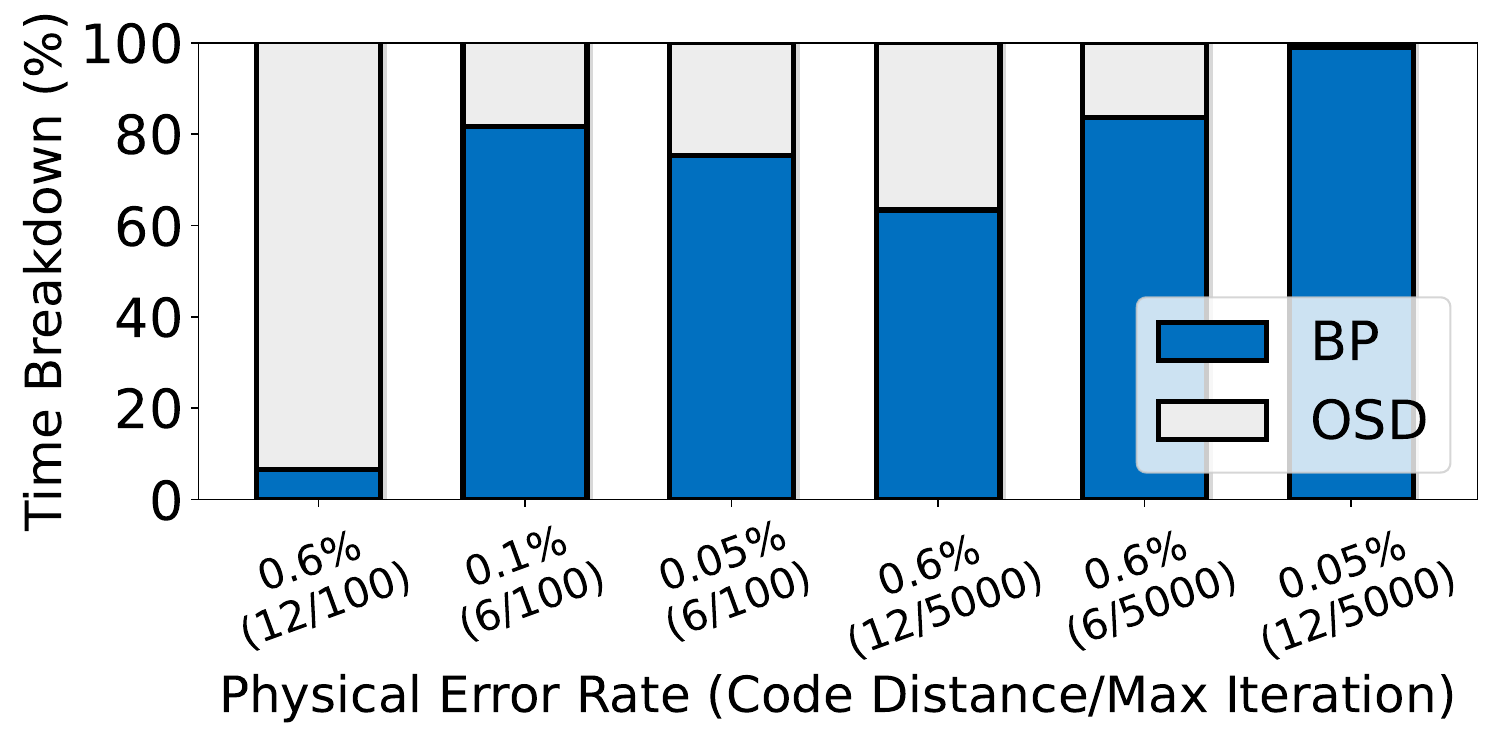}
        \caption{}
        \label{fig:bp_osd_breakdown}
    \end{subfigure}

    \begin{subfigure}{\linewidth}
        \centering
        \includegraphics[width=\linewidth]{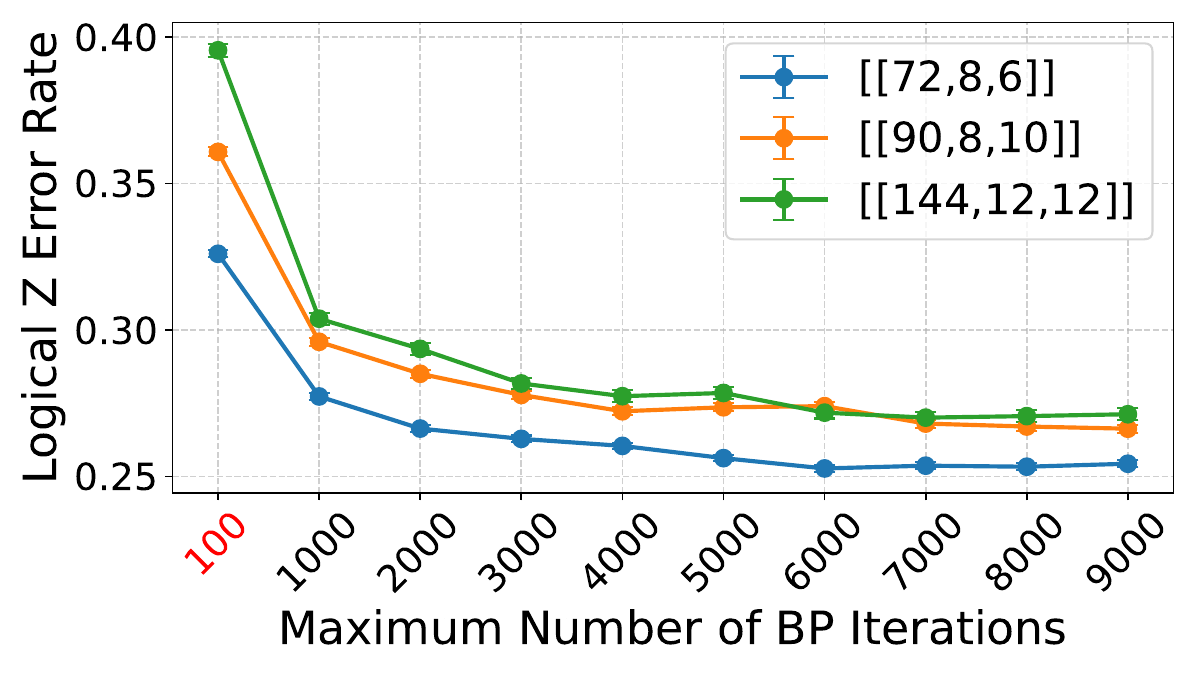}
        \caption{}
        \label{fig:Logical_Iteration}
    \end{subfigure}
    \caption{Motivation for preprocessing. (a) Decoding time breakdown of BP and OSD under varying physical error rates for BB codes with different code distances and BP iteration limits. (b) Logical
Z error rate of BB codes versus the maximum BP iteration count at a physical error rate of 0.6\%.}
    \label{fig:motivation}
\end{figure}

We first evaluated BP-OSD decoding time breakdown on two BB codes, \([[72,8,6]]\) and \([[144,12,12]]\), across some combinations of physical error rates \(\{0.6\%,\,0.1\%,\,0.05\%\}\) and maximum BP iteration limits \(\{100,\,5{,}000\}\) (the original BB code paper used \(100{,}000\) iterations~\cite{bbcode}). Here, \(0.6\%\) lies below the BB code threshold, while \(0.1\%\) and \(0.05\%\) are values expected to be achievable in near-term hardware. Figure~\ref{fig:motivation}(a) shows the breakdown of decoding time between BP and OSD. At 0.6\%, limiting BP to 100 iterations causes OSD to dominate decoding time. In contrast, at 0.1\% and 0.05\%, BP remains the dominant cost even with 100 iterations. Increasing the BP limit to 5,000 reduces the frequency of OSD executions, making BP the primary contributor to runtime across all error rates. These results demonstrate that both the BP iteration limit and physical error rate critically determine the decoding time breakdown between BP and OSD.

We further study the impact of the configured maximum number of BP iterations on logical error rate. Although we present results at a physical error rate of 0.6\%, with the decoding cycle set equal to the code distance, we observe the same trend across other physical error rates. As shown in Figure~\ref{fig:motivation}(b), aggressive early termination of BP significantly increases the logical error rate.

These results clearly demonstrate a fundamental trade-off: increasing the BP iteration limit reduces logical errors, but it also shifts most of the decoding time to BP and leads to a longer overall decoding time. Importantly, this bottleneck is not specific to BP-OSD, but is shared across a wide range of BP-based decoders. Therefore, improving the efficiency of BP offers a high-leverage optimization opportunity. Motivated by these observations, our goal is to reduce the number of required BP iterations while maintaining accuracy comparable to high-iteration configurations, thereby achieving both low logical error rates and reduced decoding latency.

\subsection{Opportunities and Challenges}

In standard BP decoding, the presence of ties (i.e., multiple error patterns having the same or nearly the same probability) can significantly slow down convergence. This is because BP cannot effectively distinguish between these candidates in the early iterations; it has to rely on many iterations to gradually accumulate weak differences through the graph, which often leads to very slow or even stalled convergence. 

Therefore, breaking ties is critical for improving BP performance. 
A natural idea is to introduce suitable local syndrome information to adjust the original channel probability vectors, thereby helping BP make distinctions more quickly and accelerating convergence. It should be noted that not all information is beneficial; only utilizing as many correct local features as possible can enhance the effectiveness of BP decoding. 
If too much misleading information is used, it may appear to break ties but does not necessarily improve convergence, as BP can become confused by such information.

Along these lines, we observe that the clique decoder~\cite{10.1145/3575693.3575733} and Pinball~\cite{pinball} for surface codes also utilize local syndrome information to improve decoding. However, their design principles cannot be directly applied to QLDPC codes due to fundamental differences in decoding structure, which introduce two key challenges.

\label{challenges}
\begin{enumerate}

\item \textbf{Compatibility with BP-based decoders.} In surface codes, the decoding problem can be reduced to an edge matching problem, which is known to be efficiently resolved partly by lightweight preprocessing, with its results directly passed to more accurate decoders such as MWPM. In contrast, QLDPC decoding involves hypergraph matching, where such information transfer is non-trivial. How to effectively propagate information from preprocessing to the BP-based decoder, therefore, becomes a key challenge.

 \item \textbf{QLDPC Decoding Complexity and Accuracy.}
Both the clique decoder and Pinball rely on identifying and directly correcting local error patterns. While this strategy is effective for surface codes, applying it to QLDPC codes leads to severe accuracy degradation due to the increased decoding complexity. We have observed that directly adopting such approaches results in a significant (at least $5\times$ in our experiments) increase in the logical error rate,  while failing to provide effective speedup.

\end{enumerate}

Thus, we need to develop fundamentally new strategies tailored to the characteristics of QLDPC codes to effectively utilize local information for accelerated decoding.

\section{Preprocessing Decoder}
\label{QPrelude}

Figure~\ref{fig:workflow} provides an overview of the preprocessing decoder workflow. The preprocessing decoder takes the raw syndrome from the extraction circuit, detects likely error events from local patterns to update the channel probability vector, and then sends both the updated probabilities and the raw syndrome into BP-based decoders for decoding.

Section~\ref{Local Syndrome Pattern Characteristics} introduces the characteristic local syndrome pattern we identified in the QLDPC code.
This pattern serves as a key indicator for detecting simple error events in the decoder.
Section~\ref{Algorithm} presents the preprocessing algorithm in detail.
Section~\ref{Hardware} describes how preprocessing decoder is implemented on an FPGA.

\subsection{Local Syndrome Pattern Characteristics}
\label{Local Syndrome Pattern Characteristics}
\begin{figure*}[ht]
  \centering
  \includegraphics[width=\textwidth]{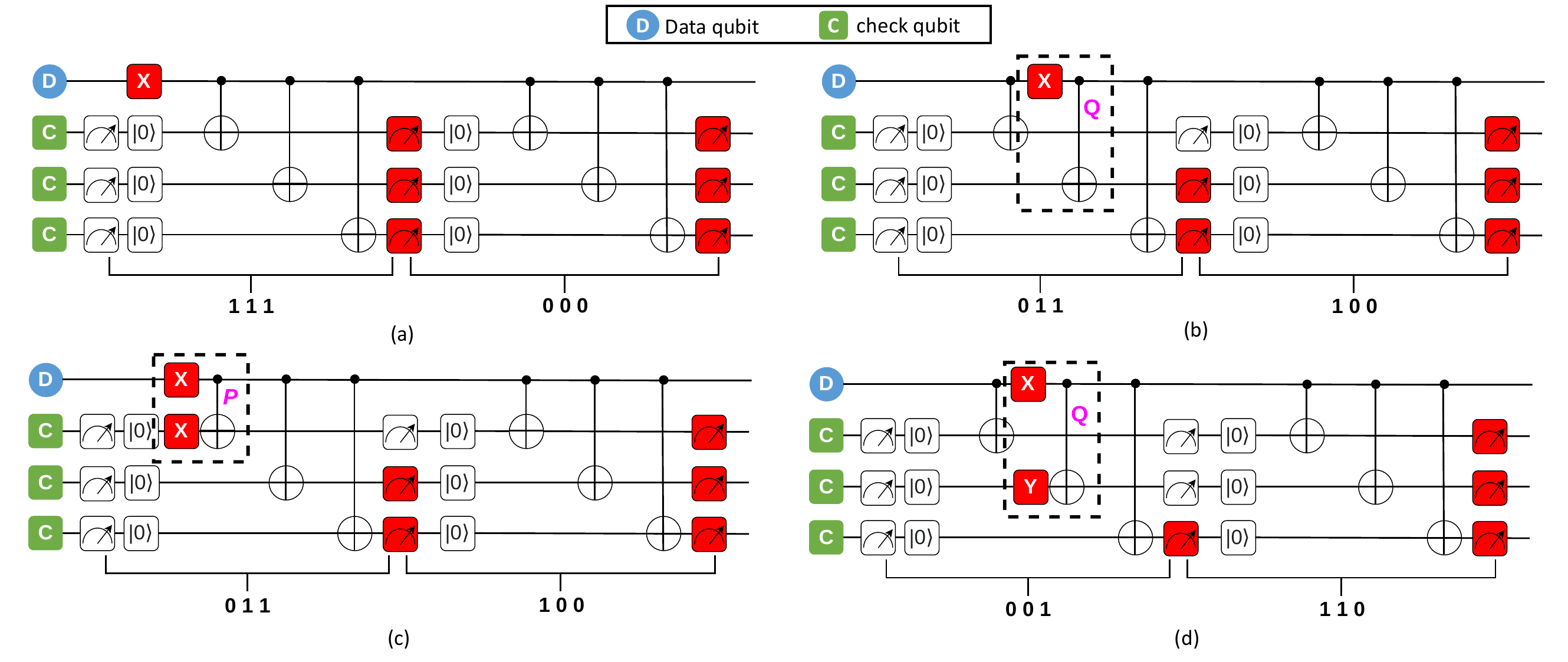}
  \caption{Four step-by-step examples of error propagation through gates in BB code syndrome extraction circuits: (a)~$X$ Error on data qubit (b)~$XI$ Error on CNOT Gate $Q$ (c)~$XX$ Error on CNOT Gate $P$ (d)~$XY$ Error on CNOT Gate $Q$.
}
  \label{fig:bb_syndrome_circuit}
\end{figure*}

In the design of preprocessing decoder, one key challenge is how to detect likely error events using only local syndrome information. To address this challenge, we analyze the syndrome patterns generated by single error events under the circuit-level noise model. These patterns can be roughly divided into three types: data errors, measurement errors, and hook errors. In preprocessing decoder, we focus on detecting single qubit data errors. Compared with hook errors, data errors are easier to identify, as hook errors can affect multiple data qubits and produce complex syndrome patterns. Compared with measurement errors, data errors are more reliable, since a single data qubit in QLDPC codes typically affects multiple check qubits, producing stronger and more distinguishable syndrome patterns. In contrast, many combinations of errors can be misidentified as measurement errors, leading to a high false detection rate.

To help understand how we detect likely error events, we use the BB code as a running example and focus on the $Z$-check side of its syndrome extraction circuit (Figure~\ref{fig:bb_syndrome_circuit}). This subcircuit involves one data qubit and three $Z$ check qubits connected by CNOT gates. Measurements are repeated over three rounds, and each check qubit is initialized to $\ket{0}$ at the beginning of each round.

We start with the simplest case. In Figure~\ref{fig:bb_syndrome_circuit}(a), suppose an $X$ error occurs on the data qubit. By comparing the outcomes of the three measurement rounds and applying XOR between consecutive rounds, we obtain detector values 111 and 000. \textbf{The XOR of detector values gives 111.} Next, we consider another case. In Figure~\ref{fig:bb_syndrome_circuit}(b), an $X$ error occurs at a different location on the data qubit. This can also be considered as an $X$ error on the control qubit of CNOT gate $P$. By comparing the measurement outcomes and applying XOR, we obtain detector values 011 and 100. \textbf{The XOR of these also gives 111.} Based on this analysis, we can systematically enumerate data-qubit errors at different locations in the BB code. We observe that the XOR of two consecutive detector values is always 111, and define this pattern as the "XOR = 111" signature.

To capture this pattern in a more general setting, we revisit Figure~\ref{fig:bb_syndrome_circuit}. In QLDPC codes, the number of check qubits connected to each data qubit may vary. Suppose a data qubit is connected to $m$ check qubits. We then describe the resulting pattern as an ''XOR = $1^m$'' signature. In BB codes, since each data qubit is connected to three Z check qubits, this corresponds to the ''XOR = 111'' pattern. This signature holds under the following conditions: (1) the code is a CSS code, where $X$- and $Z$-type syndromes are measured separately; and (2) each syndrome measurement uses an independent check qubit, with syndrome extraction performed via CNOT gates between check qubits and data qubits. These conditions are satisfied by a broad class of QLDPC codes, including BB codes and HGP codes~\cite{Kovalev_2012, Tillich_2014}.

Further analysis shows that this "XOR = $1^m$" signature is not limited to single data-qubit errors. Instead, the same signature can arise from different physical error events. Due to circuit equivalence, the error in Figure~\ref{fig:bb_syndrome_circuit}(b) is equivalent to the error shown in Figure~\ref{fig:bb_syndrome_circuit}(c), where an $XX$ error occurs on CNOT gate $Q$. In this case, we again obtain detector values 011 and 100, whose \textbf{XOR is still 111.} The same signature can also appear in more complex error patterns, such as the case shown in Figure~\ref{fig:bb_syndrome_circuit}(d).

We quantified both the fraction of single error event types detected~(event coverage) and the total probability they cover~(probability coverage) under a circuit-level noise model, as shown in Table~\ref{tab:freq}. 
For each BB code—\([[144,12,12]]\) over 12 cycles, \([[90,8,10]]\) over 10 cycles, and \([[72,8,6]]\) over 6 cycles—we used the detector error model (DEM) of Stim~\cite{Gidney2021stimfaststabilizer} to enumerate all possible single error events and then counted how many satisfy our "XOR = $1^m$" ("XOR = 111" in BB codes) signature. For example, the \([[144,12,12]]\) code has 12,060 distinct single error events in total, of which 9,183 events satisfy the signature. Therefore, the event coverage is $76.14\%$. 
To compute the probability coverage, we assume a uniform physical error rate $p$ for data qubit, measurement, idle, and initialization errors, and a rate of $p/15$ for each of the 15 CNOT gate Pauli errors. Summing the probabilities of all detected single events then gives the probability coverage values.

\begin{table}[h]
  \centering
  \caption{Error Event and Probability Coverage in BB codes}
  \label{tab:freq}

  \begin{tabular}{|c|c|c|}
    \hline
    Code & \shortstack{Event\\Coverage} & \shortstack{Probability\\Coverage} \\
    \hline
    \hline
    \([[72,8,6]]\),  \(N_c=6\)    & \(73.75\%\) & \(64.82\%\) \\
    \hline
    \([[90,8,10]]\), \(N_c=10\)   & \(75.64\%\) & \(66.90\%\) \\
    \hline
    \([[144,12,12]]\), \(N_c=12\) & \(76.14\%\) & \(67.27\%\) \\
    \hline
  \end{tabular}
\end{table}

\subsection{Heuristic Preprocessing Algorithm }
\label{Algorithm}

\begin{figure*}[t]
  \centering
  \includegraphics[width=\textwidth]{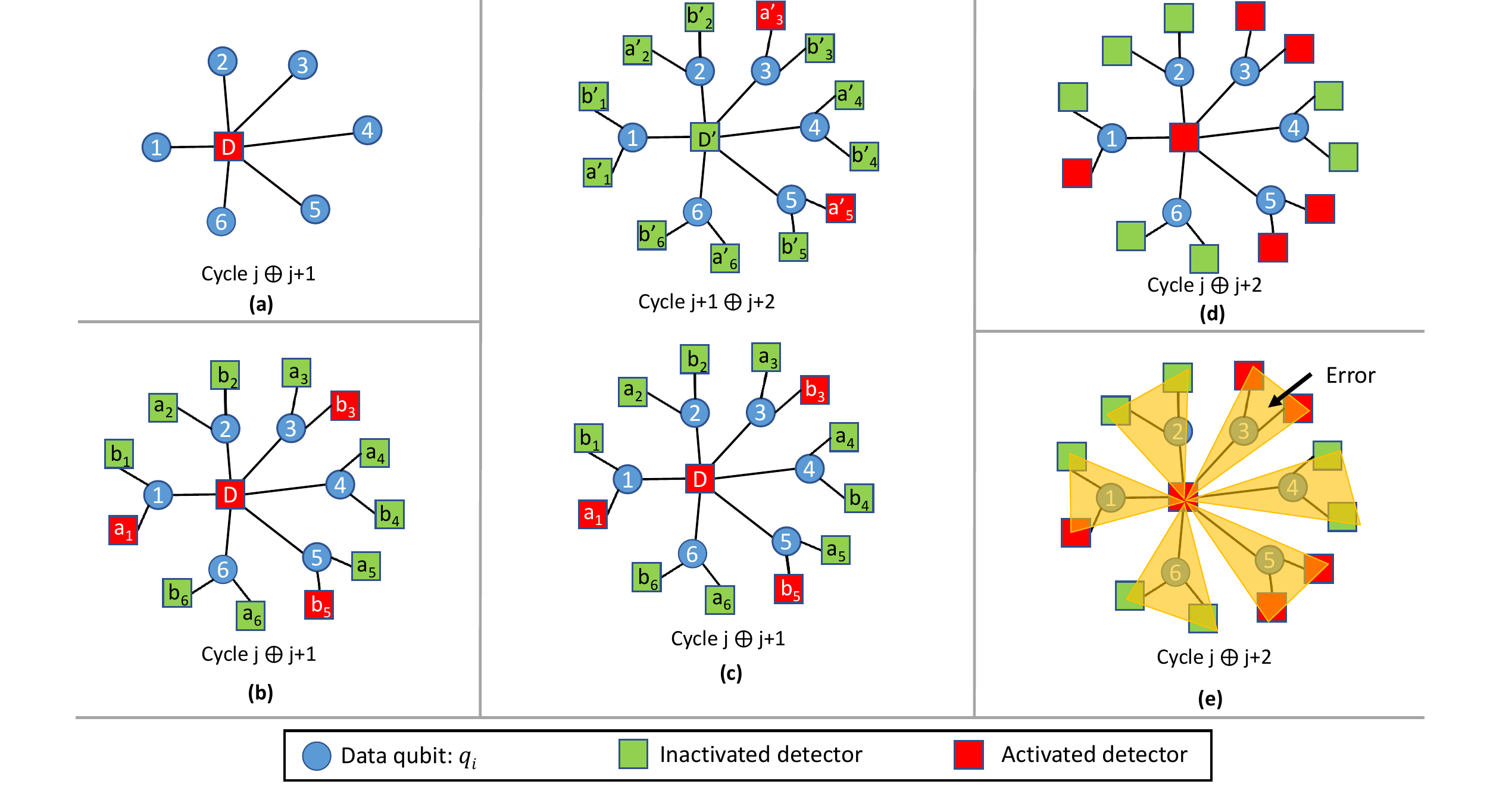}
  \caption{Step-by-step illustration of local error event detection in the preprocessing decoder. This process
analyzes XORed detectors across adjacent measurement cycles to identify error patterns. (a) Detector and connected data qubits. 
(b) Find neighboring detectors.
(c) Collect next detector values.
(d) XOR of two consecutive detector values.
(e) XOR signature probing.}
  \label{fig:work}
\end{figure*}

Based on the "XOR = $1^m$" signature, we now present the complete preprocessing decoder procedure, illustrated in Algorithm~\ref{alg:preprocess}, with a concrete BB code example shown in Figure~\ref{fig:work}. The core idea of this decoder is to scan each detector value sequentially. If a detector is activated (i.e., equals 1), we check its neighboring detectors. If the pattern satisfies the "XOR = $1^m$"  signature, it is recorded as a potential error event.

Preprocessing decoder takes two primary inputs. First is the raw syndrome $\mathit{syn}$: we extract the check qubit measurement outcomes from each cycle of the syndrome extraction circuit, then compute detector values by XORing measurements from consecutive cycles~(e.g., cycle $j\oplus j+1$). Second is the channel probability vector $channel$ as introduced in Section~\ref{BP-based Decoders}. Although the native BP-based decoder also requires additional inputs (e.g., the decoding matrix), these parameters are not modified during preprocessing and are therefore omitted from the pseudo code. The output of Algorithm~\ref{alg:preprocess} is the decoding result, indicating whether error correction succeeded. For a full description of the base QLDPC BP-based decoding algorithm, we refer the reader to IBM's original paper~\cite{bbcode}.

The decoding procedure begins by initializing two key variables (lines \ref{line:alg1_eea_def}–\ref{line:alg1_x_def}). First, we create a set $Error\_events\_A$ to store potential error events identified during preprocessing. Second, we create a copy of the raw syndrome $Syn$, denoted as $Syn\_work$, which serves as a temporary working syndrome. This allows us to preserve the original raw syndrome $Syn$ for the subsequent BP-based decoding.

We use a fixed number of iterations (lines~\ref{line:alg1_r}) to control how many times each detector is scanned.
Based on our experiments, two iterations are sufficient to detect almost all error events that match "XOR = $1^m$"  signature. This fixed-loop design is more hardware-friendly. In addition, we introduce a flag variable (line~\ref{line:alg1_init_found}) to indicate whether a valid error event has been detected during this iteration. If no error event is detected in an iteration, the loop is terminated early (line~\ref{line:alg1_endif}).

The core preprocessing routine is described in lines \ref{line:alg1_D_def}–\ref{line:alg1_endif} of Algorithm~\ref{alg:preprocess}. We illustrate each step using a BB code example shown in Figure~\ref{fig:work}:

\textbf{Find associated data qubits.} When an activated detector $D$ is identified (line~\ref{line:alg1_D_equ1}), as shown by the red square in Figure~\ref{fig:work}(a), we find the associated data qubits, denoted as $Q = \{q_1, \ldots, q_j\}$ (line~\ref{line:alg1_Q_def}). In the BB code, each check qubit is connected to six data qubits, represented by the blue circles labeled 1--6 in Figure~\ref{fig:work}(a).

\textbf{Find neighboring detectors.} We then identify all detectors connected to each data qubit $q_i$ and obtain a set of detector values $d$ (line~\ref{line:alg1_ab_def}). In the BB code, each data qubit is connected to two additional detectors. For each data qubit $q_i$, we denote these two associated detector values as $a_i$ and $b_i$, as shown in Figure~\ref{fig:work}(b).

\textbf{Collect next detector values.}
We obtain the next set of detector values, denoted by $d'$, by XORing the measurement outcomes from cycles $j+1$ and $j+2$ (line~\ref{line:alg1_abprime_def}). In the BB code example, this corresponds to detector values $a'_i$, $b'_i$, and $D'$, as shown in Figure~\ref{fig:work}(c).

\textbf{Analyze local detector values.} We compute the XOR of two consecutive detector values (cycles $j$ and $j+2$), as illustrated in Figure~\ref{fig:work}(d). Then, for the currently processed activated detector $D$, we examine each local region associated with its neighboring data qubits and evaluate the following condition (lines~\ref{line:main_loop}):
\begin{equation}
E_{q_i} = \text{all}(d \oplus d')
\end{equation}
In Figure~\ref{fig:work}(e), these regions are illustrated as six ``yellow leaves,'' each corresponding to one local region in the BB code. In this example, the condition can be expanded as:
\begin{equation}
E_{q_i} = (a_i \oplus a'_i) \land (b_i \oplus b'_i) \land (D \oplus D')
\end{equation}

Although the leaves corresponding to data qubits 3 and 5 both satisfy $E_{q_3}$ and $E_{q_5}$, in a serial software implementation, qubit 3 is detected first. Therefore, we select the region of data qubit 3 as the error event. The hardware case will be discussed in the next subsection.

\textbf{Identify error events and update control flow.} Next, in line~\ref{line:alg1_check_error_event}, we record the set of detectors that have been activated for a given data qubit (e.g., $D$, $a'_3$, and $b_3$ for data qubit 3 in Figure~\ref{fig:work}) into $d_s$. In line~\ref{line:append_event}, we use $d_s$ to search the decoding matrix to locate the corresponding error event. Once identified, the error event is added to $Error\_events\_A$ to indicate a likely error, and the corresponding activated detectors (e.g., $D$, $a'_3$, and $b_3$) are flipped to 0 in $syn\_work$ (line~\ref{line:update_syn}), without modifying the original raw syndrome.

 We then set $\mathrm{found} = \mathrm{True}$ and break out of the for-loop to proceed to the next activated detector. Because $syn\_work$ may contain new patterns that satisfy the signature after the first iteration, we perform another round of detection (line~\ref{line:alg1_r}). If no pattern matching the signature is found in the first iteration, we skip the remaining iterations (line~\ref{line:alg1_endif}).

\textbf{Channel update and BP-based decoding.} In lines ~\ref{line:alg1_channel_update} and~\ref{line:alg1_bposd}, we use the likely error events recorded in $\mathit{Error\_events\_A}$ to adjust the channel probability vector $channel$. 
Finally, we pass the updated $channel$ together with the original raw syndrome $Syn$ into BP-based decoders for decoding. The specific channel probability vector update strategies are discussed in detail in the Evaluation section (Section~\ref{Evaluation}).

\begin{algorithm}[t]
\caption{Local Syndrome-Based Preprocessing}\label{alg:preprocess}
\begin{algorithmic}[1]
\Require Raw syndrome $Syn$, channel probability vector $channel$
\Ensure  Decoding result
\State $Error\_events\_A =[]$ \label{line:alg1_eea_def}
\State $Syn\_work \gets Syn$ \label{line:alg1_x_def}
\For{$r = 1$ to $R$} \Comment{$R$ is a small fixed number, e.g., $R=2$} \label{line:alg1_r}
\State $\mathrm{found} = \mathrm{False}$ \label{line:alg1_init_found}
  \For{$D = 0$ to $|Syn\_work|-1$} \label{line:alg1_D_def}
    \If{$Syn\_work[D] = 1$} \label{line:alg1_D_equ1}
      
      \State $Q = (q_1, \ldots, q_j)$: data qubits associated with $D$  \label{line:alg1_Q_def}

      \For{$q_i \in Q$}

\State  $d$ = associated detectors for data qubit $q_i$  \label{line:alg1_ab_def}
\State  $d'$ = collect detector values in the next cycle \label{line:alg1_abprime_def}
\State  $E_{q_i} = \text{all}(d \oplus d')$ \label{line:main_loop}
  
        \If{$E_{q_i}$ is True} 
\State $d_{s}$ = activated detectors set
\label{line:alg1_check_error_event}

          \State $Error\_events\_A.append(d_s)$\label{line:append_event}
          \State $Syn\_work.update(d_s)$ \label{line:update_syn}

          \State $\mathrm{found} = \mathrm{True}$
          \State \textbf{break}

        \EndIf
      \EndFor

      \EndIf
    
  \EndFor
\If{found is False} 
        \State \textbf{break}\label{line:alg1_endif}
\EndIf 
\EndFor

\State $channel.update (Error\_events\_A)$ \label{line:alg1_channel_update}

\State $BP-Based\ Decoder(Syn,channel)$ \label{line:alg1_bposd}
\end{algorithmic}
\end{algorithm}

\subsection{Hardware Implementation}
\label{Hardware}
\begin{figure}[t]
  \centering
  \includegraphics[width=\linewidth]{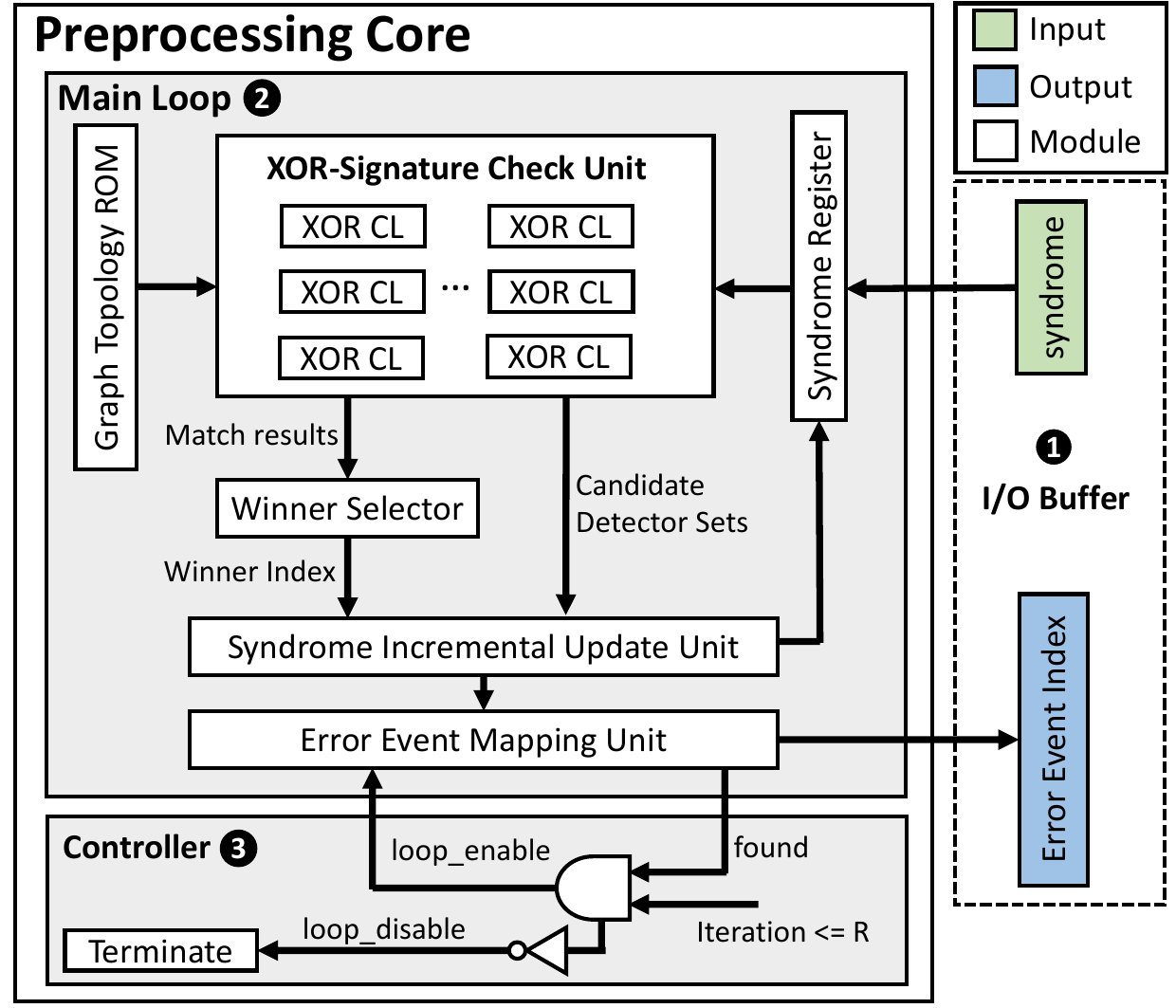}
  \caption{Hardware Architecture of the preprocessing decoder}
  \label{fig:hardware}
\end{figure}

Based on Algorithm~\ref{alg:preprocess}, we implement preprocessing decoder in hardware to demonstrate that Algorithm~\ref{alg:preprocess} is simple enough for hardware realization and can effectively reduce execution latency.

For prototyping, Algorithm~\ref{alg:preprocess} is implemented on an FPGA. This choice is motivated by the following reasons. First, a key primary advantage of FPGA-based decoders in the near term is their straightforward integration with existing quantum control systems~\cite{Battistel_2023}, which are typically implemented on FPGAs for low-level pulse control. Most of these systems already have direct access to qubit readout at the FPGA level. Therefore, this information can be efficiently transferred to the decoding FPGA through an appropriate bus interface. In addition, many representative BP-based decoders, such as BP-LSD and Relay-BP, are also designed for FPGA platforms. This allows preprocessing decoder to run alongside these decoders on the same platform, reducing system-level complexity in practical deployments.

The overall hardware architecture is shown in Figure~\ref{fig:hardware}.

\circnum{1} \textbf{Input and Output Buffer.} The raw syndrome obtained from the QLDPC measurement circuit is used as input and stored in a syndrome register. The output consists of the indices of likely error events, which are used to update the channel probability vector in the BP-based decoder.

\circnum{2} \textbf{Preprocessing Main Loop.} This is the core component of the design. We parallelize the operations corresponding to lines~\ref{line:alg1_D_equ1} to~\ref{line:main_loop} in Algorithm~\ref{alg:preprocess}, including identifying neighboring check qubits and data qubits, as well as detecting the "XOR = $1^m$" signature. 

The Graph Topology ROM stores the connectivity between check qubits and data qubits, enabling efficient lookup. The XOR-Signature Check Unit contains multiple parallel XOR check lanes~(XOR CL), where each lane corresponds to one "leaf" in Figure~\ref{fig:work}(e). The detected match results are then sent to a Winner Selector, which acts as an arbiter. When multiple candidates satisfy the signature (e.g., data qubits 3 and 5 in Figure~\ref{fig:work}(e)), the selector chooses one candidate based on a predefined priority, such as selecting the smallest index (e.g., data qubit 3). Next, the selected candidate (Winner Index) and its corresponding detector set are passed to the Syndrome Incremental Update Unit, which flips the corresponding detector values in the syndrome register, corresponding to the "activated detectors" line~\ref{line:alg1_check_error_event} in Algorithm~\ref{alg:preprocess}. In addition, the Syndrome Incremental Update Unit maps the detected signature pattern to a corresponding error event (i.e., identifying the error cycle and the data qubit location), which is then forwarded as output 

\circnum{3} \textbf{Controller.} Corresponding to line~\ref{line:alg1_r} and line~\ref{line:alg1_endif} in Algorithm~\ref{alg:preprocess}, it controls the iteration flow of preprocessing. If both conditions are satisfied, the next iteration is triggered; otherwise, the preprocessing is terminated.

\section{Numerical Evaluation}
\label{Evaluation}

In this section, we comprehensively evaluate the performance of the preprocessing decoder. We primarily evaluate our method on BB codes, a representative family of QLDPC codes. Section~\ref{Setup and Workflow} describes the evaluation procedure and experimental setup. In Section~\ref{Iteration Count and Decoding Time}, we assess the improvements in decoding latency and accuracy under a uniform circuit-level noise model across several representative baselines. To further demonstrate the robustness of preprocessing decoder, Section~\ref{Realistic Noise Model} considers a more realistic circuit-level noise model with biased error rates. In Section~\ref{hardware}, we evaluate the FPGA implementation of preprocessing decoder in terms of resource utilization and decoding latency. To demonstrate generality, we additionally validate our approach on HGP codes in Section~\ref{HGP}.

\subsection{Setup and Workflow}
\label{Setup and Workflow}
\textbf{Experimental Setup.} 
We initialize all data and check qubits, perform syndrome measurement cycles equal to the code distance, and measure the data qubits to obtain logical $Z$ syndromes. 
In our evaluation, we focus only on the $Z$ side, which is sufficient to demonstrate the decoder's performance.

We assume a circuit-level noise model with uniform physical error rate $p$:
\begin{itemize}
\item The check qubits can experience initialization errors with probability $p$.
\item The check qubits can also suffer measurement errors with probability $p$.
\item The data qubits can suffer idle errors with probability 
$p$, where for each idling gate, one of the non-identity single-qubit Pauli errors is applied with probability $p/3$.
\item For each CNOT gate, one of the 15 non-identity two-qubit Pauli errors is applied with probability $p/15$.
\end{itemize}

\begin{figure}[t]

  \centering
  \includegraphics[width=\linewidth]{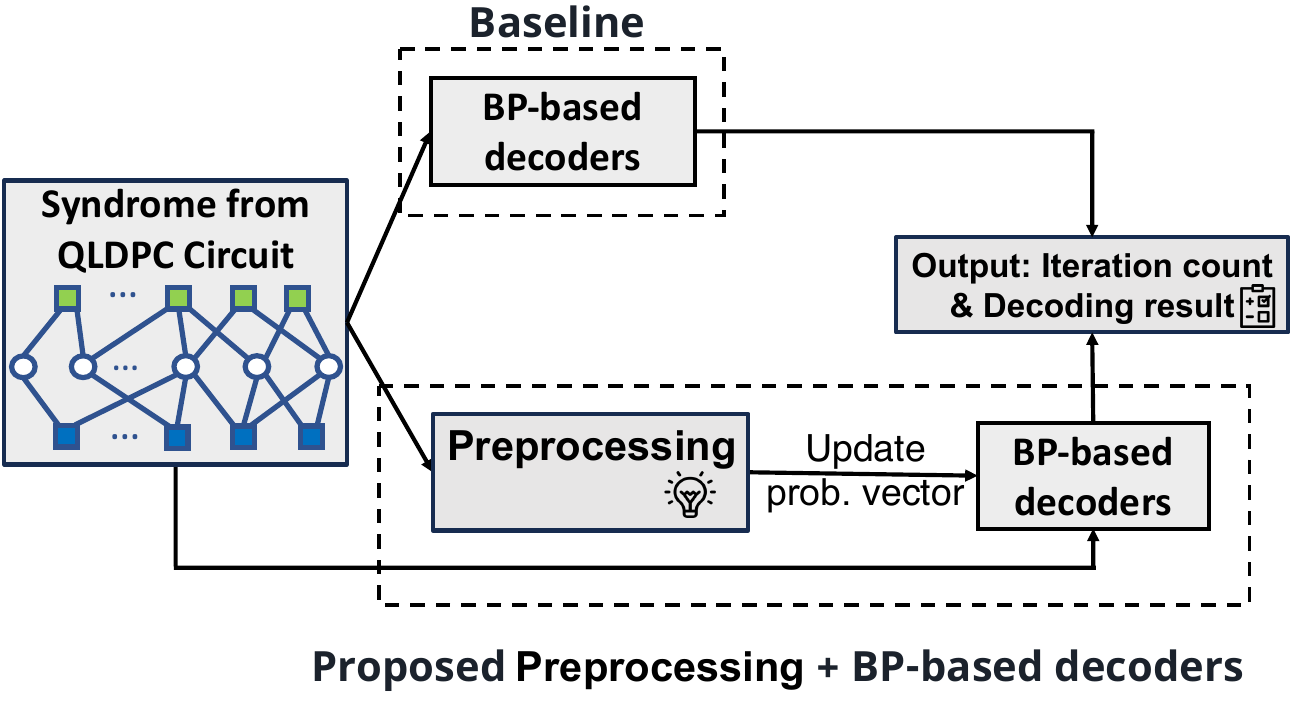}
  \caption{Evaluation Process}
  \label{fig:Evaluation}
\end{figure}

\textbf{Evaluation Workflow.} The overall procedure is shown in Figure \ref{fig:Evaluation}. For each sample, we generate a noisy QLDPC syndrome extraction circuit and record its output, which we refer to as the raw syndrome. 

The upper side of Figure \ref{fig:Evaluation} shows the baseline decoder without preprocessing. In the baseline, if the raw syndrome is zero, we skip BP-based decoding entirely, output an iteration count of zero, and verify whether the decoded logical syndrome is correct (decoding succeeds). If the raw syndrome is non-zero, we run the standard BP-based decoder, record the iteration count, and check whether decoding succeeds.
The lower side of Figure~\ref{fig:Evaluation} shows the preprocessing + BP-based decoder. To ensure a fair comparison, we reuse the same raw syndromes obtained from the baseline as input. The raw syndrome is first processed by the preprocessing step, which updates the channel probability vector. The original raw syndrome, together with the updated channel probability vector, is then passed to the BP-based decoder. Finally, we record the number of BP iterations and evaluate decoding success.

Regarding the modification of the channel probability vector, the initial probabilities for all possible single error events are obtained from detector error models (DEMs) compiled by Stim ~\cite{Gidney2021stimfaststabilizer}.

We then adjust this vector by emphasizing specific error events identified by Algorithm~\ref{alg:preprocess} ($Error\_events\_A$), scaling their corresponding probabilities with a constant factor while keeping the rest unchanged.

\textbf{Implementation.} We implement the software version of preprocessing decoder in Cython~\cite{behnel2011cython} and integrate it with Stim~\cite{Gidney2021stimfaststabilizer} and the LDPC package~\cite{Roffe_LDPC_Python_tools_2022, PhysRevResearch.2.043423}. 
Experiments in Sections~\ref{Iteration Count and Decoding Time} and~\ref{Realistic Noise Model} are conducted using the software implementation on a Mac Studio (M4 Max) with single-threaded execution. For the hardware implementation, we implement preprocessing decoder in C++ using Xilinx High-Level Synthesis (HLS) and compile it with Vitis 2024.1. We evaluate its hardware performance on a Xilinx XCU50-fsvh2104 running at 300~MHz. In Section~5.5, we evaluate preprocessing decoder in the hardware implementation combined with software Relay-BP.

\subsection{Decoding latency and accuracy }
\label{Iteration Count and Decoding Time}

\begin{figure*}[t]
  \centering
  \includegraphics[width=\textwidth]{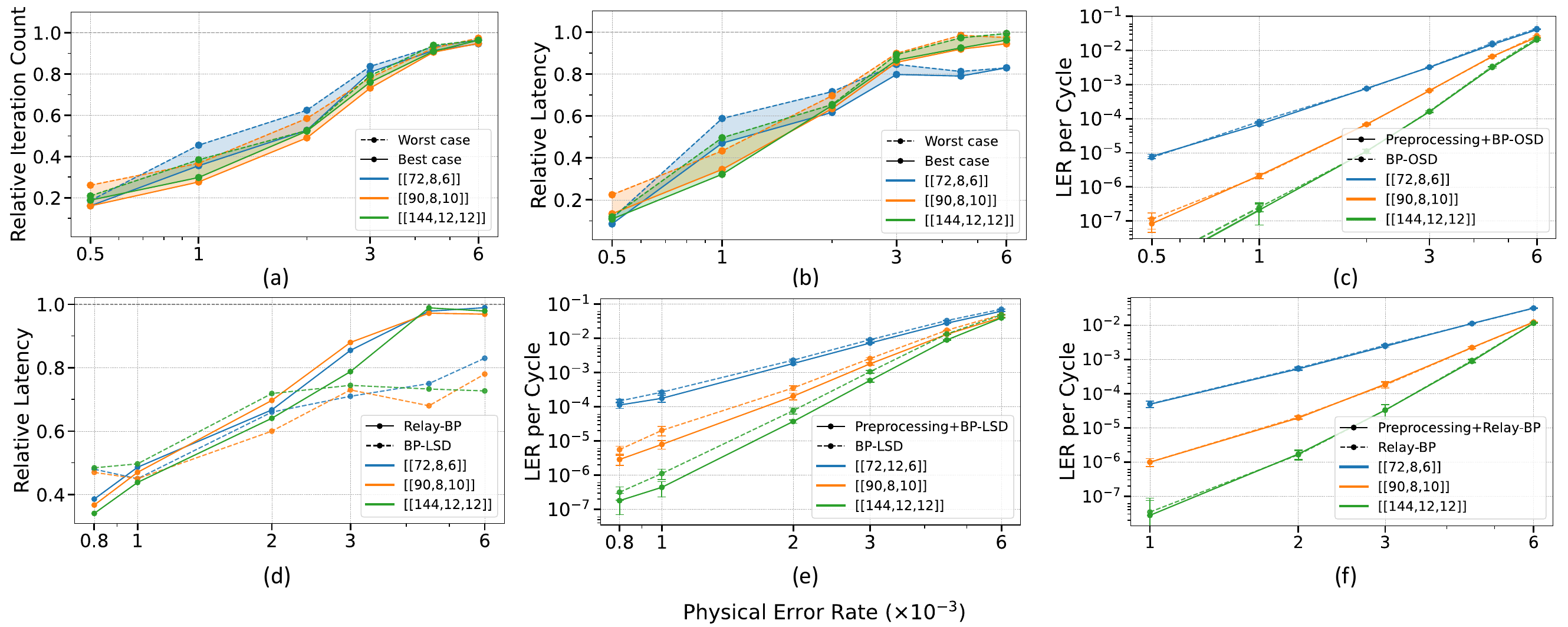}
  \caption{Impact of preprocessing on Iteration Count, Latency, and Logical Error Rate across BP-Based Decoders.(a) Relative iteration count with BP-OSD.
(b) Relative decoding latency with BP-OSD. 
(c) Logical error rate (LER) per cycle  with BP-OSD. 
(d) Relative decoding latency of Relay-BP and BP-LSD.
(e) LER per cycle with BP-LSD. 
(f) LER per cycle with Relay-BP. }
  \label{fig:main_eva}
\end{figure*}

In this subsection, we evaluate three representative BP-based decoders as baselines.
The first is BP-OSD-7, which performs sufficient BP iterations before entering OSD-7, thereby placing most of the decoding effort in the BP stage.
The second baseline is BP-LSD, which performs very limited BP iterations before entering the post-processing stage (LSD), thereby shifting more of the decoding effort to post-processing.
The third baseline is Relay-BP, which represents a BP decoder that has already been internally optimized and accelerated. This allows us to evaluate whether our method can further accelerate an already optimized BP algorithm.
Together, these three decoders represent different BP decoding strategies and are sufficient to show that our method is broadly applicable to BP-based decoders.

\textbf{BP-OSD-7:}
To evaluate the performance of our method under this type of decoder, we follow the procedure described in the motivation section to estimate the number of iterations required for sufficient convergence under different physical error rates. Accordingly, the maximum iteration counts for p = 0.05\%, 0.1\%, 0.2\%, 0.3\%, 0.45\%, and 0.6\% are set to 100, 300, 500, 1,000, 3,000, and 5,000, respectively. 

We adjust the channel probability vector by scaling the prior probabilities of the corresponding error events. We test scaling factors of 2, 4, and 6. The result with the worst-performing scaling factor is reported as the worst case (dashed line) in Figure~\ref{fig:main_eva}(a) and (b), while the result with the best-performing scaling factor is reported as the best case (solid line). We evaluate the relative iteration count and relative decoding time for BB codes with d=6,10, and 12 under different physical error rates. The relative iteration count is defined as
\begin{equation}
\label{eq:rel_iter}
\text{Relative iteration count} =
\frac{\text{Preprocessing + BP-OSD iteration count}}
{\text{BP-OSD iteration count}}.
\end{equation}
Similarly, the relative decoding time is defined as the ratio between the total runtime of the preprocessing+baseline decoder and that of the baseline decoder. 

We observe that at lower physical error rates, the number of iterations is significantly reduced, leading to substantial improvements in decoding speed — about 10× faster. This is because at low error rates, errors are sparsely distributed, allowing the preprocessing step to effectively utilize local syndrome patterns for error detection. However, as the physical error rate increases, error patterns become more complex, making it more difficult for preprocessing alone to resolve errors. Consequently, the performance improvement diminishes at higher error rates. For the logical error rate (LER), Figure~\ref{fig:main_eva}(c) shows that preprocessing decoder has almost the same performance as BP-OSD. All reported logical error rate results are obtained using scaling factors 2 and 4, as they achieve similar accuracy. In contrast, a scaling factor of 6 is too aggressive in this setting and is therefore excluded. Overall, preprocessing decoder reduces decoding time while maintaining the same accuracy as BP-OSD.

\textbf{BP-LSD:}
We configure the decoder according to the original BP-LSD paper, where the maximum number of BP iterations is set to 30. We use a scaling factor of 4. We observe a similar trend in decoding time reduction. As shown by the dashed line in Figure~\ref{fig:main_eva}(d), the decoding time is still significantly reduced. In this setting, we do not report the reduction in BP iterations, since the maximum number of BP iterations is fixed to 30. As a result, the decoder almost always reaches this upper bound, making iteration reduction less meaningful to compare.

In terms of decoding accuracy, Figure~\ref{fig:main_eva}(e) shows that our method achieves much better performance than the baseline. Specifically, our method reduces the logical error rate by approximately 1.3×–1.8× across different code distances, with larger improvements observed for higher distance codes. This is because BP in this setting cannot fully converge due to the limited number of iterations. Our preprocessing step effectively helps BP move closer to convergence, improving the quality of its output. As a result, the LSD stage can make better decisions. This result also supports our observation in the motivation section: the number of BP iterations has a direct impact on the final decoding accuracy. With more effective BP updates (or better intermediate results), the decoding accuracy can be improved. Overall, for this type of decoder, preprocessing decoder achieves both speedup and accuracy improvement.

\textbf{Relay-BP:}
We configure the decoder according to the original paper and set R = 301 for the Relay-BP-1 experiments. We also use a scaling factor of 4. As shown by the solid line in Figure~\ref{fig:main_eva}(d), we observe a similar trend in decoding speed improvement compared to other baselines. In addition, Figure~\ref{fig:main_eva}(f) shows that the decoding accuracy is well preserved. Therefore, our preprocessing method can further accelerate the decoder reported by IBM as one of the fastest and most accurate decoders~\cite{ibm_relay_bp_blog}, while maintaining its accuracy.

\subsection{Evaluation under Biased Noise Models}
\label{Realistic Noise Model}

In practical quantum hardware systems, different quantum operations are implemented using different physical mechanisms, and their error rates are therefore not uniform. As a result, compared to a uniform noise model that assigns the same error rate to all operations, realistic noise often exhibits a clear imbalance among different types of error events. This motivates the use of biased noise models for a more faithful evaluation of preprocessing decoder performance.

\begin{figure*}[ht]

  \includegraphics[width=\linewidth]{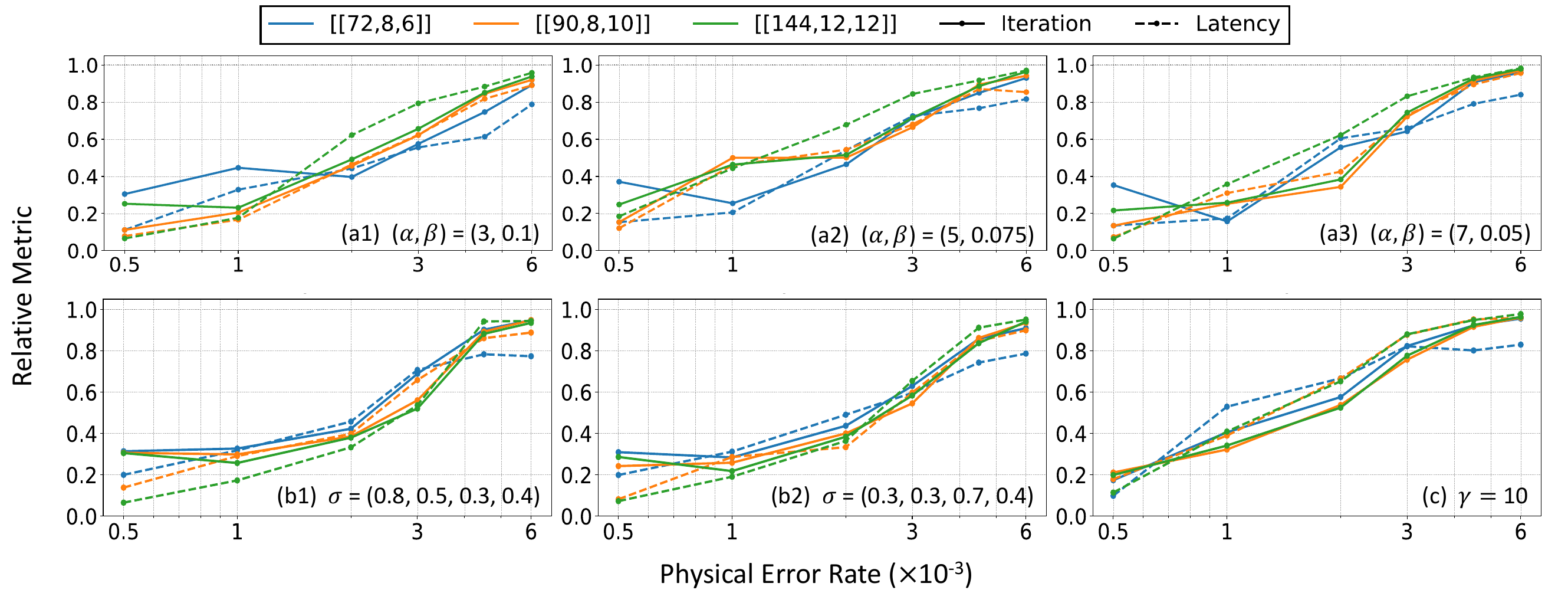}

  \caption{Preprocessing under biased noise models:
(a) bias between gate types for three $(\alpha,\beta)$ settings, 
(b) bias between qubits for two $\sigma$ configurations, 
and (c) Pauli-type bias.}
\label{fig:biased}
\end{figure*}

To evaluate the robustness of preprocessing decoder under biased noise, we conduct three groups of independent experiments based on BP-OSD, each targeting a different source of noise bias. The channel probability update factor is fixed to four ($\times 4$). For all three bias models, we report the relative iteration count and latency, as shown in Figure~\ref{fig:biased}.

\textbf{Bias between gate types:}
We fix the two-qubit gate error rate to be $p$, set the measurement error rate on check qubits to be $\alpha p$, and set the single-qubit gate error rate to be $\beta p$.
By adjusting the parameters $\alpha$ and $\beta$, we control the strength of the bias among different operation types. 
We evaluate three $(\alpha,\beta)$ settings.
The setting, $(\alpha,\beta) =(3,0.1)$, corresponds to a relatively moderate bias and closely reflects the current ratio among single-qubit gate, two-qubit gate, and measurement errors observed in Google~\cite{Willow} and IBM devices~\cite{IBMquantumcomputers}.
The setting, $(\alpha,\beta) =(5,0.075)$, corresponds to a more aggressive bias regime, and  $(\alpha,\beta) =(7,0.05)$, is used as a stress-test case. 
 The results are shown in Figure~\ref{fig:biased} (a1--a3). Preprocessing decoder shows clear acceleration under all these settings, indicating that it is robust to bias between different gate types.

\textbf{Bias between qubits:}
We next conduct bias between different qubits. To model uncertainty during circuit execution, we apply random perturbations to the error rates:
$p \mapsto  p\times(1+\delta), \quad \delta \sim \mathcal{N}(0, \sigma^2).$
The values of $\sigma$ are chosen based on reported data for Google’s Willow chip and public device information~\cite{Willow,GoogleWillowSpecSheet2024}. We evaluate two sets of qubit-dependent noise variations across initialization, two-qubit, measurement, and idle operations, with
$\sigma = (0.8, 0.5, 0.3, 0.4)$ for the first set and
$\sigma = (0.3, 0.3, 0.7, 0.4)$ for the second set, respectively. The results are shown in Figure~\ref{fig:biased} (b1--b2). Even with non-uniform noise across qubits, preprocessing decoder still provides strong acceleration.

\textbf{Bias between Pauli error types:}
Finally, we consider bias between different Pauli error types. In many qubit systems, $Z$ errors are more likely than $X$ or $Y$ errors. We use a heuristic Pauli error bias model. 
Two-qubit Pauli errors that do not contain $Z$ are assigned probability $p/15$, 
while errors that contain at least one $Z$ are assigned probability $(p/15)\times \gamma$. We choose $\gamma = 10$ in our experiments. The results in Figure~\ref{fig:biased}(c) show that preprocessing decoder remains effective under Pauli-type bias.

In addition, we confirm that the logical error rate results under these biased noise models are consistent with the conclusions in Section~\ref{Iteration Count and Decoding Time}. Overall, these experiments show that preprocessing decoder is robust to noise bias and noise uncertainty, and works well under more realistic noise conditions.

\subsection{Hardware Utilization and Latency}
\label{hardware}

\begin{table}[h]
  \centering
  \caption{FPGA Resource Utilization for QLDPC Codes}
  \label{tab:Hardware}

  \begin{tabular}{|c|c|c|}
    \hline
    Code & LUT & FF \\
    \hline
    \hline
     \([[72,8,6]]\),  \(N_c=6\)    & 5867 (\(0.67\%\)) & 6149 (\(0.35\%\)) \\
   
    \hline
    \([[90,8,10]]\), \(N_c=10\)   & 7329 (\(0.84\%\))& 7632 (\(0.44\%\)) \\
    \hline
    \([[144,12,12]]\), \(N_c=12\) & 10153 (\(1.16\%\)) & 9527 (\(0.54\%\)) \\
    \hline
  \end{tabular}
\end{table}

Table~\ref{tab:Hardware} shows the FPGA resource utilization of preprocessing decoder for different BB codes under the setting $N_c = d$, where $N_c$ denotes the number of syndrome measurement cycles and is set equal to the code distance. The results show that preprocessing decoder is highly lightweight. For the $[[144,12,12]]$ code with $d=12$ cycles, it consumes less than 2\% of LUTs and less than 1\% of FFs on a Xilinx XCU50 FPGA. Based on this, we estimate that the design can scale to support on the order of $10^4$ data qubits.

\begin{figure*}[ht]

  \includegraphics[width=\linewidth]{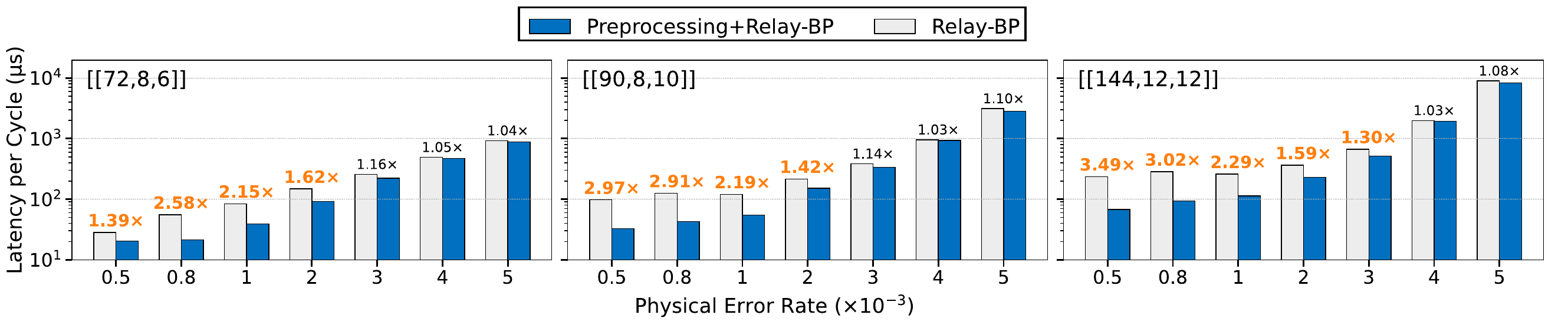}
  \caption{Latency Comparison per Syndrome Measurement Cycle under Relay-BP with and without preprocessing}
  \label{fig:latency_per_cycle}
\end{figure*}

Figure~\ref{fig:latency_per_cycle} compares the per syndrome measurement cycle latency of preprocessing decoder (hardware) combined with Relay-BP (software) against Relay-BP (software)  alone. We observe that as the code distance increases, the performance improvement becomes larger at low physical error rates, indicating strong scalability. For the $[[72,8,6]]$, $[[90,8,10]]$, and $[[144,12,12]]$ codes, the worst-case latency per syndrome measurement cycle of preprocessing is 251~ns, 332~ns, and 480~ns, respectively. On average, the preprocessing decoding latency is less than 1\% of the total decoding time.

\subsection{Evaluation on HGP Codes}
\label{HGP}

\begin{figure}[t]
  \centering
  \includegraphics[width=\linewidth]{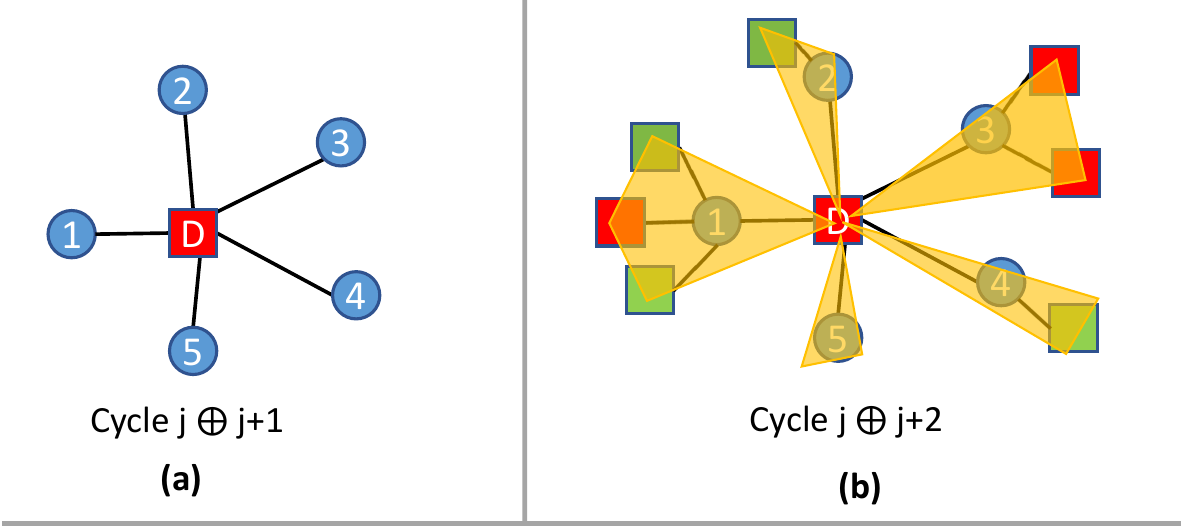}
  \caption{XOR signature detection in HGP codes. (a) Detector and connected data qubits.  (b) XOR signature probing.
}
  \label{fig:adaptation}
\end{figure}

Hypergraph product (HGP) codes are a class of QLDPC codes constructed from two classical linear codes via the hypergraph product. To illustrate how our approach applies to such codes, Figure~\ref{fig:adaptation} illustrates a portion of an HGP code. For a flipped detector $D$ connected to five data qubits, we apply the same preprocessing procedure as in Figure~\ref{fig:work}(b)–(e), resulting in Figure~\ref{fig:adaptation}(b). When probing the leaves for possible error events, the XOR condition is adapted according to the connectivity of each data qubit; for example, data qubit~1 requires "XOR = 1111", while data qubit~2 requires only "XOR = 11".

\begin{figure}[t]
  \centering
  \includegraphics[width=\linewidth]{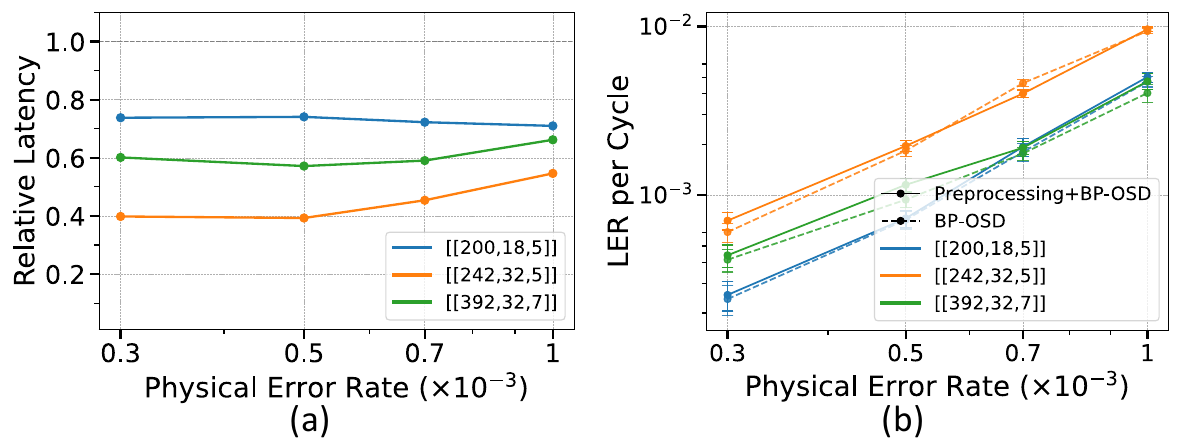}
  \caption{Comparison of decoding performance between preprocessing + BP-OSD and baseline BP-OSD under HGP codes. (a) Relative decoding latency (b) Logical error rate (per cycle)}
  \label{fig:hgp_eva}
\end{figure}

 We generate HGP codes based on the methodology in~\cite{HGP_con} and evaluate preprocessing decoder combined with BP-OSD on three HGP codes ([[200,18,5]], [[242,32,5]], [[392,32,7]]). The relative decoding latency and logical error rate per syndrome measurement cycle are shown in Figure~\ref{fig:hgp_eva}. Our results show that preprocessing decoder, with the channel probability vector scaling factor ($\times$1.5), also achieves speedups on HGP codes while maintaining decoding accuracy.

 However, we observe several interesting differences compared to BB codes. First, in BB codes, the speedup exhibits a relatively linear trend with respect to the physical error rate (Figure~\ref{fig:main_eva}), whereas for HGP codes, the trend is less regular. This difference is likely due to the high structural symmetry of BB codes, which leads to more stable behavior as the physical error rate varies. In contrast, HGP codes lack such symmetry, and the trend appears less regular. Second, we observe that HGP codes are more sensitive to the parameters used in updating the channel probability vector. Although more aggressive scaling (e.g., $\times4$, $\times6$) can significantly reduce decoding time, they may incur a slight loss in accuracy. However, with a more appropriate choice of scaling factor, more aggressive scaling may further accelerate decoding while maintaining accuracy. In HGP codes, the number of check qubits connected to each data qubit varies. As a result, the confidence of an error event detected by an ``XOR = $1^m$'' pattern depends on the value of $m$. Larger $m$ indicates higher confidence in the detected error event, suggesting that more aggressive update strategies may be more suitable in such cases. A more principled modeling of this adaptive scaling approach is left as future work.

\section{Conclusion}

In this paper, we propose a lightweight local syndrome-based preprocessing method that enhances BP-based decoding for general QLDPC codes. We develop a complete decoding pipeline and implement preprocessing in both software and FPGA-based hardware. We evaluate its performance across diverse noise settings and a range of representative BP-based decoding baselines. The results show that preprocessing significantly reduces decoding time, and in some cases, even reduces the logical error rate. These improvements bring QLDPC decoding one step closer to practical deployment.

\begin{acks}

This work was partly supported by the Feasibility Study on the Future HPCI, JST Moonshot R\&D Grant Numbers JPMJMS2061, JPMJMS2067, and JPMJMS226C; JST CREST Grant Numbers JPMJCR23I4 and JPMJCR24I4; MEXT Q-LEAP Grant Numbers JPMXS0120319794 and JPMXS0118068682; and JSPS KAKENHI Grant Numbers JP22H05000, JP24K02915, JP25K03094, and JP25K21176; and RIKEN Special Postdoctoral Researcher Program. This material is based upon work supported by the U.S. Department of Energy, Office of Science, Office of Advanced Scientific Computing Research, Accelerated Research in Quantum Computing under Award Number DE-SC0025633. 
\end{acks}


\bibliographystyle{ACM-Reference-Format}
\bibliography{sample-base}

@book{nielsen2002quantum,
  author    = {Michael A. Nielsen and Isaac L. Chuang},
  title     = {Quantum Computation and Quantum Information},
  year      = {2002},
  publisher = {Cambridge University Press},
  isbn      = {978-0521635035}
}

@article{quantumc,
	author = {Feynman, Richard P. },
	journal = {International Journal of Theoretical Physics},
	number = {6},
	pages = {467--488},
	title = {Simulating physics with computers},
	volume = {21},
	year = {1982}}

@ARTICLE{shor1997,
	author = {Shor, Peter W.},
	journal = {SIAM Journal on Computing},
	number = {5},
	pages = {1484-1509},
	title = {Polynomial-Time Algorithms for Prime Factorization and Discrete Logarithms on a Quantum Computer},
	volume = {26},
	year = {1997}}

@inproceedings{10.1145/237814.237866,
author = {Grover, Lov K.},
title = {A fast quantum mechanical algorithm for database search},
year = {1996},
isbn = {0897917855},
publisher = {Association for Computing Machinery},
address = {New York, NY, USA},
url = {https://doi.org/10.1145/237814.237866},
doi = {10.1145/237814.237866},
booktitle = {Proceedings of the Twenty-Eighth Annual ACM Symposium on Theory of Computing},
pages = {212–219},
numpages = {8},
location = {Philadelphia, Pennsylvania, USA},
series = {STOC '96}
}

@article{Preskill2018quantumcomputingin,
  doi = {10.22331/q-2018-08-06-79},
  url = {https://doi.org/10.22331/q-2018-08-06-79},
  title = {Quantum {C}omputing in the {NISQ} era and beyond},
  author = {Preskill, John},
  journal = {{Quantum}},
  issn = {2521-327X},
  publisher = {{Verein zur F{\"{o}}rderung des Open Access Publizierens in den Quantenwissenschaften}},
  volume = {2},
  pages = {79},
  month = aug,
  year = {2018}
}

@ARTICLE{5437474,
  author={Zhang, Zhengya and Anantharam, Venkat and Wainwright, Martin J. and Nikolic, Borivoje},
  journal={IEEE Journal of Solid-State Circuits}, 
  title={An Efficient 10GBASE-T Ethernet LDPC Decoder Design With Low Error Floors}, 
  year={2010},
  volume={45},
  number={4},
  pages={843-855},
  doi={10.1109/JSSC.2010.2042255}}

@misc{bravyi1998quantumcodeslatticeboundary,
      title={Quantum codes on a lattice with boundary}, 
      author={S. B. Bravyi and A. Yu. Kitaev},
      year={1998},
      eprint={quant-ph/9811052},
      archivePrefix={arXiv},
      primaryClass={quant-ph},
      url={https://arxiv.org/abs/quant-ph/9811052}, 
}

@article{kitaev1997quantum,
  author    = {A. Yu. Kitaev},
  title     = {Quantum Computations: Algorithms and Error Correction},
  journal   = {Russian Mathematical Surveys},
  volume    = {52},
  number    = {6},
  pages     = {1191--1249},
  year      = {1997},
  month     = {dec},
  doi       = {10.1070/RM1997v052n06ABEH002155},
  url       = {https://doi.org/10.1070/RM1997v052n06ABEH002155}
}

@article{PhysRevLett.98.190504,
  title = {Fault-Tolerant Quantum Computation with High Threshold in Two Dimensions},
  author = {Raussendorf, Robert and Harrington, Jim},
  journal = {Phys. Rev. Lett.},
  volume = {98},
  issue = {19},
  pages = {190504},
  numpages = {4},
  year = {2007},
  month = {May},
  publisher = {American Physical Society},
  doi = {10.1103/PhysRevLett.98.190504},
  url = {https://link.aps.org/doi/10.1103/PhysRevLett.98.190504}
}

@article{Kitaev_2003,
   title={Fault-tolerant quantum computation by anyons},
   volume={303},
   ISSN={0003-4916},
   url={http://dx.doi.org/10.1016/S0003-4916(02)00018-0},
   DOI={10.1016/s0003-4916(02)00018-0},
   number={1},
   journal={Annals of Physics},
   publisher={Elsevier BV},
   author={Kitaev, A.Yu.},
   year={2003},
   month=jan, pages={2–30} }

@article{gottesman2014faulttolerantquantumcomputationconstant,
author = {Gottesman, Daniel},
title = {Fault-tolerant quantum computation with constant overhead},
year = {2014},
issue_date = {November 2014},
publisher = {Rinton Press, Incorporated},
address = {Paramus, NJ},
volume = {14},
number = {15–16},
issn = {1533-7146},
journal = {Quantum Info. Comput.},
month = nov,
pages = {1338–1372},
numpages = {35}
}

@article{PRXQuantum.2.040101,
  title = {Quantum Low-Density Parity-Check Codes},
  author = {Breuckmann, Nikolas P. and Eberhardt, Jens Niklas},
  journal = {PRX Quantum},
  volume = {2},
  issue = {4},
  pages = {040101},
  numpages = {19},
  year = {2021},
  month = {Oct},
  publisher = {American Physical Society},
  doi = {10.1103/PRXQuantum.2.040101},
  url = {https://link.aps.org/doi/10.1103/PRXQuantum.2.040101}
}

@ARTICLE{1057683,
  author={Gallager, R.},
  journal={IRE Transactions on Information Theory}, 
  title={Low-density parity-check codes}, 
  year={1962},
  volume={8},
  number={1},
  pages={21-28},
  doi={10.1109/TIT.1962.1057683}}

@ARTICLE{8316763,
  author={Richardson, Tom and Kudekar, Shrinivas},
  journal={IEEE Communications Magazine}, 
  title={Design of Low-Density Parity Check Codes for 5G New Radio}, 
  year={2018},
  volume={56},
  number={3},
  pages={28-34},
  doi={10.1109/MCOM.2018.1700839}}

@inproceedings{10.1145/3564246.3585169,
author = {Gu, Shouzhen and Pattison, Christopher A. and Tang, Eugene},
title = {An Efficient Decoder for a Linear Distance Quantum LDPC Code},
year = {2023},
isbn = {9781450399135},
publisher = {Association for Computing Machinery},
address = {New York, NY, USA},
url = {https://doi.org/10.1145/3564246.3585169},
doi = {10.1145/3564246.3585169},
booktitle = {Proceedings of the 55th Annual ACM Symposium on Theory of Computing},
pages = {919–932},
numpages = {14},
location = {Orlando, FL, USA},
series = {STOC 2023}
}

@inproceedings{10.1145/3564246.3585101,
author = {Dinur, Irit and Hsieh, Min-Hsiu and Lin, Ting-Chun and Vidick, Thomas},
title = {Good Quantum LDPC Codes with Linear Time Decoders},
year = {2023},
isbn = {9781450399135},
publisher = {Association for Computing Machinery},
address = {New York, NY, USA},
url = {https://doi.org/10.1145/3564246.3585101},
doi = {10.1145/3564246.3585101},
pages = {905–918},
numpages = {14},
location = {Orlando, FL, USA},
series = {STOC 2023}
}

@article{article111,
author = {Leverrier, Anthony and Zémor, Gilles},
year = {2024},
month = {05},
pages = {},
title = {Efficient decoding up to a constant fraction of the code length for asymptotically good quantum codes},
journal = {ACM Transactions on Algorithms},
doi = {10.1145/3663763}
}

@article{Single-Shot,
	author = {Gu, Shouzhen and Tang, Eugene and Caha, Libor and Choe, Shin Ho and He, Zhiyang and Kubica, Aleksander},
	journal = {Communications in Mathematical Physics},
	number = {3},
	pages = {85},
	title = {Single-Shot Decoding of Good Quantum LDPC Codes},
	volume = {405},
	year = {2024}}

@article{Panteleev_2021,
   title={Degenerate Quantum LDPC Codes With Good Finite Length Performance},
   volume={5},
   ISSN={2521-327X},
   url={http://dx.doi.org/10.22331/q-2021-11-22-585},
   DOI={10.22331/q-2021-11-22-585},
   journal={Quantum},
   publisher={Verein zur Forderung des Open Access Publizierens in den Quantenwissenschaften},
   author={Panteleev, Pavel and Kalachev, Gleb},
   year={2021},
   month=nov, pages={585} }

@misc{morris2024absorbingsetsquantumldpc,
      title={Absorbing Sets in Quantum LDPC Codes}, 
      author={Kirsten D. Morris and Tefjol Pllaha and Christine A. Kelley},
      year={2024},
      eprint={2307.14532},
      archivePrefix={arXiv},
      primaryClass={cs.IT},
      url={https://arxiv.org/abs/2307.14532}, 
      }

@article{PhysRevResearch.2.043423,
  title = {Decoding across the quantum low-density parity-check code landscape},
  author = {Roffe, Joschka and White, David R. and Burton, Simon and Campbell, Earl},
  journal = {Phys. Rev. Res.},
  volume = {2},
  issue = {4},
  pages = {043423},
  numpages = {13},
  year = {2020},
  month = {Dec},
  publisher = {American Physical Society},
  doi = {10.1103/PhysRevResearch.2.043423},
  url = {https://link.aps.org/doi/10.1103/PhysRevResearch.2.043423}
}

@article{terhal2015quantum,
  title={Quantum error correction for quantum memories},
  author={Terhal, Barbara M},
  journal={Reviews of Modern Physics},
  volume={87},
  number={2},
  pages={307},
  year={2015},
  publisher={APS}
}

@inproceedings{holmes2020nisq,
author = {Holmes, Adam and Jokar, Mohammad Reza and Pasandi, Ghasem and Ding, Yongshan and Pedram, Massoud and Chong, Frederic T.},
title = {{NISQ}+: Boosting Quantum Computing Power by Approximating Quantum Error Correction},
year = {2020},
isbn = {9781728146614},
doi = {10.1109/ISCA45697.2020.00053},
booktitle = {Proceedings of the ACM/IEEE 47th Annual International Symposium on Computer Architecture},
pages = {556–569},
numpages = {14},
}

@article{battistel2023real,
doi = {10.1088/2399-1984/aceba6},
year = {2023},
month = {aug},
publisher = {IOP Publishing},
volume = {7},
number = {3},
pages = {032003},
author={Battistel, Francesco and Chamberland, Christopher and Johar, Kauser and Overwater, Ramon WJ and Sebastiano, Fabio and Skoric, Luka and Ueno, Yosuke and Usman, Muhammad},
title = {Real-time decoding for fault-tolerant quantum computing: progress, challenges and outlook},
journal = {Nano Futures},
}

@article{PhysRevA.54.1098,
  title = {Good quantum error-correcting codes exist},
  author = {Calderbank, A. R. and Shor, Peter W.},
  journal = {Phys. Rev. A},
  volume = {54},
  issue = {2},
  pages = {1098--1105},
  numpages = {0},
  year = {1996},
  month = {Aug},
  publisher = {American Physical Society},
  doi = {10.1103/PhysRevA.54.1098},
  url = {https://link.aps.org/doi/10.1103/PhysRevA.54.1098}
}

@article{PhysRevLett.77.793,
  title = {Error Correcting Codes in Quantum Theory},
  author = {Steane, A. M.},
  journal = {Phys. Rev. Lett.},
  volume = {77},
  issue = {5},
  pages = {793--797},
  numpages = {0},
  year = {1996},
  month = {Jul},
  publisher = {American Physical Society},
  doi = {10.1103/PhysRevLett.77.793},
  url = {https://link.aps.org/doi/10.1103/PhysRevLett.77.793}
}

@inproceedings{Kovalev_2012,
   title={Improved quantum hypergraph-product LDPC codes},
   url={http://dx.doi.org/10.1109/ISIT.2012.6284206},
   DOI={10.1109/isit.2012.6284206},
   booktitle={2012 IEEE International Symposium on Information Theory Proceedings},
   publisher={IEEE},
   author={Kovalev, Alexey A. and Pryadko, Leonid P.},
   year={2012},
   month=jul, pages={348–352} }

@article{Tillich_2014,
   title={Quantum LDPC Codes With Positive Rate and Minimum Distance Proportional to the Square Root of the Blocklength},
   volume={60},
   ISSN={1557-9654},
   url={http://dx.doi.org/10.1109/TIT.2013.2292061},
   DOI={10.1109/tit.2013.2292061},
   number={2},
   journal={IEEE Transactions on Information Theory},
   publisher={Institute of Electrical and Electronics Engineers (IEEE)},
   author={Tillich, Jean-Pierre and Zemor, Gilles},
   year={2014},
   month=feb, pages={1193–1202} }

@misc{li2024transformarbitrarygoodquantum,
      title={Transform Arbitrary Good Quantum LDPC Codes into Good Geometrically Local Codes in Any Dimension}, 
      author={Xingjian Li and Ting-Chun Lin and Min-Hsiu Hsieh},
      year={2024},
      eprint={2408.01769},
      archivePrefix={arXiv},
      primaryClass={quant-ph},
      url={https://arxiv.org/abs/2408.01769}, 
}

@misc{koukoulekidis2024smallquantumcodesalgebraic,
      title={Small Quantum Codes from Algebraic Extensions of Generalized Bicycle Codes}, 
      author={Nikolaos Koukoulekidis and Fedor Šimkovic IV and Martin Leib and Francisco Revson Fernandes Pereira},
      year={2024},
      eprint={2401.07583},
      archivePrefix={arXiv},
      primaryClass={quant-ph},
      url={https://arxiv.org/abs/2401.07583}, 
}

@article{bbcode,
	author = {Bravyi, Sergey and Cross, Andrew W. and Gambetta, Jay M. and Maslov, Dmitri and Rall, Patrick and Yoder, Theodore J.},
	journal = {Nature},
	number = {8005},
	pages = {778--782},
	title = {High-threshold and low-overhead fault-tolerant quantum memory},
	volume = {627},
	year = {2024}}

@article{Fowler_2012,
   title={Surface codes: Towards practical large-scale quantum computation},
   volume={86},
   ISSN={1094-1622},
   url={http://dx.doi.org/10.1103/PhysRevA.86.032324},
   DOI={10.1103/physreva.86.032324},
   number={3},
   journal={Physical Review A},
   publisher={American Physical Society (APS)},
   author={Fowler, Austin G. and Mariantoni, Matteo and Martinis, John M. and Cleland, Andrew N.},
   year={2012},
   month=sep }

@article{Higgott_2025,
   title={Sparse Blossom: correcting a million errors per core second with minimum-weight matching},
   volume={9},
   ISSN={2521-327X},
   url={http://dx.doi.org/10.22331/q-2025-01-20-1600},
   DOI={10.22331/q-2025-01-20-1600},
   journal={Quantum},
   publisher={Verein zur Forderung des Open Access Publizierens in den Quantenwissenschaften},
   author={Higgott, Oscar and Gidney, Craig},
   year={2025},
   month=jan, pages={1600} }

@INPROCEEDINGS {pinball,
author = { Knapen, Alexander and Tao, Guanchen and Mack, Jacob and Bruno, Tomas and Saligane, Mehdi and Sylvester, Dennis and Zhang, Qirui and Ravi, Gokul Subramanian },
booktitle = { 2026 IEEE International Symposium on High Performance Computer Architecture (HPCA) },
title = {{ Pinball: A Cryogenic Predecoder for Quantum Error Correction Decoding Under Circuit-Level Noise }},
year = {2026},
volume = {},
ISSN = {},
pages = {1-17},
doi = {10.1109/HPCA68181.2026.11408464},
url = {https://doi.ieeecomputersociety.org/10.1109/HPCA68181.2026.11408464},
publisher = {IEEE Computer Society},
address = {Los Alamitos, CA, USA},
month =Feb}

@inproceedings{10.1145/3620666.3651339,
author = {Alavisamani, Narges and Vittal, Suhas and Ayanzadeh, Ramin and Das, Poulami and Qureshi, Moinuddin},
title = {Promatch: Extending the Reach of Real-Time Quantum Error Correction with Adaptive Predecoding},
year = {2024},
isbn = {9798400703867},
publisher = {Association for Computing Machinery},
address = {New York, NY, USA},
url = {https://doi.org/10.1145/3620666.3651339},
doi = {10.1145/3620666.3651339},
booktitle = {Proceedings of the 29th ACM International Conference on Architectural Support for Programming Languages and Operating Systems, Volume 3},
pages = {818–833},
numpages = {16},
location = {La Jolla, CA, USA},
series = {ASPLOS '24}
}

@article{Smith_2023,
   title={Local Predecoder to Reduce the Bandwidth and Latency of Quantum Error Correction},
   volume={19},
   ISSN={2331-7019},
   url={http://dx.doi.org/10.1103/PhysRevApplied.19.034050},
   DOI={10.1103/physrevapplied.19.034050},
   number={3},
   journal={Physical Review Applied},
   publisher={American Physical Society (APS)},
   author={Smith, Samuel C. and Brown, Benjamin J. and Bartlett, Stephen D.},
   year={2023},
   month=mar }

@article{Chamberland_2023,
   title={Techniques for combining fast local decoders with global decoders under circuit-level noise},
   volume={8},
   ISSN={2058-9565},
   url={http://dx.doi.org/10.1088/2058-9565/ace64d},
   DOI={10.1088/2058-9565/ace64d},
   number={4},
   journal={Quantum Science and Technology},
   publisher={IOP Publishing},
   author={Chamberland, Christopher and Goncalves, Luis and Sivarajah, Prasahnt and Peterson, Eric and Grimberg, Sebastian},
   year={2023},
   month=jul, pages={045011} }

@misc{relay-bp,
      title={Improved belief propagation is sufficient for real-time decoding of quantum memory}, 
      author={Tristan Müller and Thomas Alexander and Michael E. Beverland and Markus Bühler and Blake R. Johnson and Thilo Maurer and Drew Vandeth},
      year={2025},
      eprint={2506.01779},
      archivePrefix={arXiv},
      primaryClass={quant-ph},
      url={https://arxiv.org/abs/2506.01779}, 
}

@article{hillmann2024localizedstatisticsdecodingparallel,
	author = {Hillmann, Timo and Berent, Lucas and Quintavalle, Armanda O. and Eisert, Jens and Wille, Robert and Roffe, Joschka},
	journal = {Nature Communications},
	number = {1},
	pages = {8214},
	title = {Localized statistics decoding for quantum low-density parity-check codes},
	volume = {16},
	year = {2025}}

@INPROCEEDINGS{10821172,
  author={Wolanski, Stasiu and Barber, Ben},
  booktitle={2024 IEEE International Conference on Quantum Computing and Engineering (QCE)}, 
  title={Introducing Ambiguity Clustering: An Accurate and Efficient Decoder for qLDPC Codes}, 
  year={2024},
  volume={02},
  number={},
  pages={402-403},
  doi={10.1109/QCE60285.2024.10326}}

@misc{iolius2024almostlineartimedecodingalgorithm,
      title={An almost-linear time decoding algorithm for quantum LDPC codes under circuit-level noise}, 
      author={Antonio deMarti iOlius and Imanol Etxezarreta Martinez and Joschka Roffe and Josu Etxezarreta Martinez},
      year={2024},
      eprint={2409.01440},
      archivePrefix={arXiv},
      primaryClass={quant-ph},
      url={https://arxiv.org/abs/2409.01440}, 
}

@misc{gong2024lowlatencyiterativedecodingqldpc,
      title={Toward Low-latency Iterative Decoding of QLDPC Codes Under Circuit-Level Noise}, 
      author={Anqi Gong and Sebastian Cammerer and Joseph M. Renes},
      year={2024},
      eprint={2403.18901},
      archivePrefix={arXiv},
      primaryClass={quant-ph},
      url={https://arxiv.org/abs/2403.18901}, 
}

@misc{yin2024symbreakmitigatingquantumdegeneracy,
      title={SymBreak: Mitigating Quantum Degeneracy Issues in QLDPC Code Decoders by Breaking Symmetry}, 
      author={Keyi Yin and Xiang Fang and Jixuan Ruan and Hezi Zhang and Dean Tullsen and Andrew Sornborger and Chenxu Liu and Ang Li and Travis Humble and Yufei Ding},
      year={2024},
      eprint={2412.02885},
      archivePrefix={arXiv},
      primaryClass={quant-ph},
      url={https://arxiv.org/abs/2412.02885}, 
}

@inbook{10.1145/3725843.3756084,
author = {Zhou, Kaiwen and Lu, Liqiang and Xiang, Debin and Tao, Chenning and Wu, Anbang and Leng, Jingwen and Liu, Fangxin and Chen, Mingshuai and Yin, Jianwei},
title = {Vegapunk: Accurate and Fast Decoding for Quantum LDPC Codes with Online Hierarchical Algorithm and Sparse Accelerator},
year = {2025},
isbn = {9798400715730},
publisher = {Association for Computing Machinery},
address = {New York, NY, USA},
url = {https://doi.org/10.1145/3725843.3756084},
booktitle = {Proceedings of the 58th IEEE/ACM International Symposium on Microarchitecture},
pages = {719–732},
numpages = {14}
}

@misc{wang2025fullyparallelizedbpdecoding,
      title={Fully Parallelized BP Decoding for Quantum LDPC Codes Can Outperform BP-OSD}, 
      author={Ming Wang and Ang Li and Frank Mueller},
      year={2025},
      eprint={2507.00254},
      archivePrefix={arXiv},
      primaryClass={quant-ph},
      url={https://arxiv.org/abs/2507.00254}, 
}

@software{cudaqx,
  author       = {NVIDIA Corporation, CUDA-QX Development Team},
  title        = {CUDA-QX},
  year         = {2025},
  publisher    = {NVIDIA},
  address      = {Santa Clara, CA},
  url          = {https://github.com/NVIDIA/cudaqx},
  note         = {Accessed: 2025-08-17}
}

@misc{blue2025machinelearningdecodingcircuitlevel,
      title={Machine Learning Decoding of Circuit-Level Noise for Bivariate Bicycle Codes}, 
      author={John Blue and Harshil Avlani and Zhiyang He and Liu Ziyin and Isaac L. Chuang},
      year={2025},
      eprint={2504.13043},
      archivePrefix={arXiv},
      primaryClass={quant-ph},
      url={https://arxiv.org/abs/2504.13043}, 
}

@article{ninkovic2024decodingquantumldpccodes,
  title={Decoding Quantum LDPC Codes Using Graph Neural Networks},
  author={Vukan Ninkovic and Ognjen Kundacina and Dejan Vukobratovi{\'c} and Christian H{\"a}ger and Alexandre Graell i Amat},
  journal={GLOBECOM 2024 - 2024 IEEE Global Communications Conference},
  year={2024},
  pages={3479-3484},
  url={https://api.semanticscholar.org/CorpusID:271843131}
}

@article{maan2024machinelearningmessagepassingscalable,
	author = {Maan, Arshpreet Singh and Paler, Alexandru},
	journal = {npj Quantum Information},
	number = {1},
	pages = {78},
	title = {Machine learning message-passing for the scalable decoding of QLDPC codes},
	volume = {11},
	year = {2025}}

@article{PhysRevLett.122.200501,
  title = {Neural Belief-Propagation Decoders for Quantum Error-Correcting Codes},
  author = {Liu, Ye-Hua and Poulin, David},
  journal = {Phys. Rev. Lett.},
  volume = {122},
  issue = {20},
  pages = {200501},
  numpages = {6},
  year = {2019},
  month = {May},
  publisher = {American Physical Society},
  doi = {10.1103/PhysRevLett.122.200501},
  url = {https://link.aps.org/doi/10.1103/PhysRevLett.122.200501}
}

@article{Gidney2021stimfaststabilizer,
  doi = {10.22331/q-2021-07-06-497},
  url = {https://doi.org/10.22331/q-2021-07-06-497},
  title = {Stim: a fast stabilizer circuit simulator},
  author = {Gidney, Craig},
  journal = {{Quantum}},
  issn = {2521-327X},
  publisher = {{Verein zur F{\"{o}}rderung des Open Access Publizierens in den Quantenwissenschaften}},
  volume = {5},
  pages = {497},
  month = jul,
  year = {2021}
}

@software{Roffe_LDPC_Python_tools_2022,
author = {Roffe, Joschka},
title = {{LDPC: Python tools for low density parity check codes}},
url = {https://pypi.org/project/ldpc/},
year = {2022}
}

@article{Willow,
	author = {Acharya, Rajeev and Abanin, Dmitry A. and Aghababaie-Beni, Laleh and Aleiner, Igor and Andersen, Trond I. and Ansmann, Markus and Arute, Frank and Arya, Kunal and Asfaw, Abraham and Astrakhantsev, Nikita and Atalaya, Juan and Babbush, Ryan and Bacon, Dave and Ballard, Brian and Bardin, Joseph C. and Bausch, Johannes and Bengtsson, Andreas and Bilmes, Alexander and Blackwell, Sam and Boixo, Sergio and Bortoli, Gina and Bourassa, Alexandre and Bovaird, Jenna and Brill, Leon and Broughton, Michael and Browne, David A. and Buchea, Brett and Buckley, Bob B. and Buell, David A. and Burger, Tim and Burkett, Brian and Bushnell, Nicholas and Cabrera, Anthony and Campero, Juan and Chang, Hung-Shen and Chen, Yu and Chen, Zijun and Chiaro, Ben and Chik, Desmond and Chou, Charina and Claes, Jahan and Cleland, Agnetta Y. and Cogan, Josh and Collins, Roberto and Conner, Paul and Courtney, William and Crook, Alexander L. and Curtin, Ben and Das, Sayan and Davies, Alex and De Lorenzo, Laura and Debroy, Dripto M. and Demura, Sean and Devoret, Michel and Di Paolo, Agustin and Donohoe, Paul and Drozdov, Ilya and Dunsworth, Andrew and Earle, Clint and Edlich, Thomas and Eickbusch, Alec and Elbag, Aviv Moshe and Elzouka, Mahmoud and Erickson, Catherine and Faoro, Lara and Farhi, Edward and Ferreira, Vinicius S. and Burgos, Leslie Flores and Forati, Ebrahim and Fowler, Austin G. and Foxen, Brooks and Ganjam, Suhas and Garcia, Gonzalo and Gasca, Robert and Genois, {\'E}lie and Giang, William and Gidney, Craig and Gilboa, Dar and Gosula, Raja and Dau, Alejandro Grajales and Graumann, Dietrich and Greene, Alex and Gross, Jonathan A. and Habegger, Steve and Hall, John and Hamilton, Michael C. and Hansen, Monica and Harrigan, Matthew P. and Harrington, Sean D. and Heras, Francisco J. H. and Heslin, Stephen and Heu, Paula and Higgott, Oscar and Hill, Gordon and Hilton, Jeremy and Holland, George and Hong, Sabrina and Huang, Hsin-Yuan and Huff, Ashley and Huggins, William J. and Ioffe, Lev B. and Isakov, Sergei V. and Iveland, Justin and Jeffrey, Evan and Jiang, Zhang and Jones, Cody and Jordan, Stephen and Joshi, Chaitali and Juhas, Pavol and Kafri, Dvir and Kang, Hui and Karamlou, Amir H. and Kechedzhi, Kostyantyn and Kelly, Julian and Khaire, Trupti and Khattar, Tanuj and Khezri, Mostafa and Kim, Seon and Klimov, Paul V. and Klots, Andrey R. and Kobrin, Bryce and Kohli, Pushmeet and Korotkov, Alexander N. and Kostritsa, Fedor and Kothari, Robin and Kozlovskii, Borislav and Kreikebaum, John Mark and Kurilovich, Vladislav D. and Lacroix, Nathan and Landhuis, David and Lange-Dei, Tiano and Langley, Brandon W. and Laptev, Pavel and Lau, Kim-Ming and Le Guevel, Lo{\"\i}ck and Ledford, Justin and Lee, Joonho and Lee, Kenny and Lensky, Yuri D. and Leon, Shannon and Lester, Brian J. and Li, Wing Yan and Li, Yin and Lill, Alexander T. and Liu, Wayne and Livingston, William P. and Locharla, Aditya and Lucero, Erik and Lundahl, Daniel and Lunt, Aaron and Madhuk, Sid and Malone, Fionn D. and Maloney, Ashley and Mandr{\`a}, Salvatore and Manyika, James and Martin, Leigh S. and Martin, Orion and Martin, Steven and Maxfield, Cameron and McClean, Jarrod R. and McEwen, Matt and Meeks, Seneca and Megrant, Anthony and Mi, Xiao and Miao, Kevin C. and Mieszala, Amanda and Molavi, Reza and Molina, Sebastian and Montazeri, Shirin and Morvan, Alexis and Movassagh, Ramis and Mruczkiewicz, Wojciech and Naaman, Ofer and Neeley, Matthew and Neill, Charles and Nersisyan, Ani and Neven, Hartmut and Newman, Michael and Ng, Jiun How and Nguyen, Anthony and Nguyen, Murray and Ni, Chia-Hung and Niu, Murphy Yuezhen and O'Brien, Thomas E. and Oliver, William D. and Opremcak, Alex and Ottosson, Kristoffer and Petukhov, Andre and Pizzuto, Alex and Platt, John and Potter, Rebecca and Pritchard, Orion and Pryadko, Leonid P. and Quintana, Chris and Ramachandran, Ganesh and Reagor, Matthew J. and Redding, John and Rhodes, David M. and Roberts, Gabrielle and Rosenberg, Eliott and Rosenfeld, Emma and Roushan, Pedram and Rubin, Nicholas C. and Saei, Negar and Sank, Daniel and Sankaragomathi, Kannan and Satzinger, Kevin J. and Schurkus, Henry F. and Schuster, Christopher and Senior, Andrew W. and Shearn, Michael J. and Shorter, Aaron and Shutty, Noah and Shvarts, Vladimir and Singh, Shraddha and Sivak, Volodymyr and Skruzny, Jindra and Small, Spencer and Smelyanskiy, Vadim and Smith, W. Clarke and Somma, Rolando D. and Springer, Sofia and Sterling, George and Strain, Doug and Suchard, Jordan and Szasz, Aaron and Sztein, Alex and Thor, Douglas and Torres, Alfredo and Torunbalci, M. Mert and Vaishnav, Abeer and Vargas, Justin and Vdovichev, Sergey and Vidal, Guifre and Villalonga, Benjamin and Heidweiller, Catherine Vollgraff and Waltman, Steven and Wang, Shannon X. and Ware, Brayden and Weber, Kate and Weidel, Travis and White, Theodore and Wong, Kristi and Woo, Bryan W. K. and Xing, Cheng and Yao, Z. Jamie and Yeh, Ping and Ying, Bicheng and Yoo, Juhwan and Yosri, Noureldin and Young, Grayson and Zalcman, Adam and Zhang, Yaxing and Zhu, Ningfeng and Zobrist, Nicholas and Google Quantum AI and Collaborators},
	journal = {Nature},
	number = {8052},
	pages = {920--926},
	title = {Quantum error correction below the surface code threshold},
	volume = {638},
	year = {2025}}

@article{IBMquantumcomputers,
	author = {AbuGhanem, Muhammad},
	journal = {The Journal of Supercomputing},
	number = {5},
	pages = {687},
	title = {IBM quantum computers: evolution, performance, and future directions},
	volume = {81},
	year = {2025}}

@misc{GoogleWillowSpecSheet2024,
  author       = {{Google Quantum AI}},
  title        = {Willow Spec Sheet},
  howpublished = {Technical report / specification sheet},
  year         = {2024},
  month        = dec,
  note         = {Published Dec 9, 2024. Accessed: 2025-12-26},
  url          = {https://quantumai.google/static/site-assets/downloads/willow-spec-sheet.pdf}
}

@article{Battistel_2023,
	author = {Battistel, F and Chamberland, C and Johar, K and Overwater, R W J and Sebastiano, F and Skoric, L and Ueno, Y and Usman, M},
	journal = {Nano Futures},
	month = {aug},
	number = {3},
	pages = {032003},
	title = {Real-time decoding for fault-tolerant quantum computing: progress, challenges and outlook},
	volume = {7},
	year = {2023}}

@inproceedings{10.1145/3575693.3575733,
author = {Ravi, Gokul Subramanian and Baker, Jonathan M. and Fayyazi, Arash and Lin, Sophia Fuhui and Javadi-Abhari, Ali and Pedram, Massoud and Chong, Frederic T.},
title = {Better Than Worst-Case Decoding for Quantum Error Correction},
year = {2023},
isbn = {9781450399166},
publisher = {Association for Computing Machinery},
address = {New York, NY, USA},
url = {https://doi.org/10.1145/3575693.3575733},
doi = {10.1145/3575693.3575733},
booktitle = {Proceedings of the 28th ACM International Conference on Architectural Support for Programming Languages and Operating Systems, Volume 2},
pages = {88–102},
numpages = {15},
location = {Vancouver, BC, Canada},
series = {ASPLOS 2023}
}

@misc{ibm_relay_bp_blog,
  author = {{IBM Quantum}},
  title = {Relay-BP error correction decoder},
  year = {2023},
  howpublished = {\url{https://www.ibm.com/quantum/blog/relay-bp-error-correction-decoder}},
  note = {Accessed: 2026-03-29}
}

@article{behnel2011cython,
  title={Cython: The best of both worlds},
  author={Behnel, Stefan and Bradshaw, Robert and Citro, Craig and Dalcin, Lisandro and Seljebotn, Dag Sverre and Smith, Kurt},
  journal={Computing in Science \& Engineering},
  volume={13},
  number={2},
  pages={31--39},
  year={2011},
  publisher={IEEE}
}

@article{HGP_con,
	author = {Pecorari, Laura and Jandura, Sven and Brennen, Gavin K. and Pupillo, Guido},
	journal = {Nature Communications},
	number = {1},
	pages = {1111},
	title = {High-rate quantum LDPC codes for long-range-connected neutral atom registers},
	volume = {16},
	year = {2025}}

\end{document}